\documentclass[11pt,twoside]{article}
\usepackage{amsfonts}
\usepackage{times}
\usepackage[vcentermath,enableskew]{youngtab}
\usepackage{graphicx}
\DeclareMathAlphabet{\mathsfsl}{OT1}{cmss}{m}{sl}
\DeclareMathAlphabet{\mathscr}{U}{eus}{m}{n}
\DeclareMathAlphabet{\matheur}{U}{eur}{m}{n}
\SetMathAlphabet{\matheur}{bold}{U}{eur}{b}{n}
\newcommand{\nn}{\nonumber \\}
\newcommand{\beq}{\begin{equation}} 
\newcommand{\eeq}{\end{equation}}
\newcommand{\bea}{\begin{eqnarray}}
\newcommand{\eea}{\end{eqnarray}}
\newcommand{\Ref}[1]{(\ref{#1})}

\newcommand{\pa}{\partial}
\newcommand{\tr}{{\rm tr}\,}

\newcommand{\A}{\alpha}
\newcommand{\B}{\beta}
\newcommand{\oB}{\overline{\beta}}
\newcommand{\C}{\gamma}
\newcommand{\GA}{\Gamma}
\newcommand{\D}{\delta}

\newcommand{\E}{\varepsilon}

\newcommand{\N}{\nu}

\newcommand{\SI}{\sigma}

\newcommand{\OM}{\omega}
\newcommand{\OO}{\Omega}

\newcommand{\HC}{\mathcal{H}}
\newcommand{\VC}{\mathcal{V}}

\newcommand{\DC}{\mathcal{D}}
\newcommand{\BC}{\mathcal{B}}
\newcommand{\IC}{\mathcal{I}}
\newcommand{\LC}{\mathcal{L}}
\newcommand{\NC}{\mathcal{N}}
\newcommand{\OC}{\mathcal{O}}

\newcommand{\R}{\mathcal{R}}

\newcommand{\gl}{{\mathfrak{sl}}_{10}}
\newcommand{\DD}{{\; / \;}}

\newcommand{\Rn}{{\bf R}}
\newcommand{\Zn}{{\bf Z}}

\begin{document}
\def\draft{\pagestyle{draft}\thispagestyle{draft}
\global\def\draftcontrol{1}}
\global\def\draftcontrol{0}
\arraycolsep3pt

\thispagestyle{empty}
\begin{flushright} hep-th/0504153
                   \\  AEI-2005-051
\end{flushright}
\vspace*{1.0cm}
\begin{center}
 {\LARGE \sc Higher Order M Theory Corrections \\[1ex]
             and the Kac-Moody Algebra $E_{10}$
  }\\
 \vspace*{1cm}
 {\sl
     Thibault Damour\footnotemark[1] and
     Hermann Nicolai\footnotemark[2] \\
 \vspace*{6mm}
     \footnotemark[1]
     Institut des Hautes Etudes Scientifiques\\
     35, Route de Chartres, F-91440 Bures-sur-Yvette, France\\
 \vspace*{3mm}
     \footnotemark[2]
     Max-Planck-Institut f\"ur Gravitationsphysik\\
     Albert-Einstein-Institut \\
     M\"uhlenberg 1, D-14476 Potsdam, Germany} \\
 \vspace*{1cm}
\begin{minipage}{11cm}\footnotesize
\textbf{Abstract:}
It has been conjectured that the classical dynamics of M theory
is equivalent to a null geodesic motion in the infinite-dimensional
coset space $E_{10}/K(E_{10})$, where $K(E_{10})$ is the maximal
compact subgroup of the hyperbolic Kac-Moody group $E_{10}$. We here
provide further evidence for this conjecture by showing that the
leading higher order corrections, quartic in the curvature and related
three-form dependent terms, correspond to {\em negative imaginary roots}
of $E_{10}$. The conjecture entails certain predictions for which 
higher order corrections are allowed: in particular corrections 
of type $R^M (DF)^N$ are compatible with $E_{10}$ only for $M+N=3k+1$. 
Furthermore, the leading parts of the $ R^4, R^7,\cdots $ terms are 
predicted to be associated with {\it singlets} under the $\gl$
decomposition of $E_{10}$. Although singlets are extremely rare among 
the altogether $4\,400\,752\,653$ representations of $\gl$ appearing 
in $E_{10}$ up to level $\ell \leq 28$, there are indeed singlets
at levels $\ell=10$ and $\ell =20$ which do match with the $R^4$ and 
the expected $R^7$ corrections. Our analysis indicates a far more 
complicated behavior of the theory near the cosmological singularity 
than suggested by the standard homogeneous {\it ans\"atze}.

\end{minipage}
\end{center}
\setcounter{footnote}{0}

\section{Introduction}
The analysis \`a la Belinskii, Khalatnikov, Lifshitz (BKL) \cite{BKL}
of generic cosmological solutions of $D=11$ supergravity \cite{CrJuSche78}
in the vicinity of a spacelike singularity has revealed a connection 
with billiard motion in the fundamental Weyl chamber of $E_{10}$ 
(implying chaotic oscillations of the metric near the singularity)
\cite{DaHe01,DaHeJuNi01,DaHeNi03}. We recall that $E_{10}$ is a 
 rank-$10$ infinite-dimensional hyperbolic Kac-Moody
algebra \cite{Kac}\footnote{For simplicity of notation, $E_{10}$ denotes 
both the {\it group}, and its associated Lie {\it algebra}.} whose root 
lattice is the canonical hyperbolic extension of the root lattice 
of the largest exceptional finite-dimensional Lie algebra $E_8$,
and the unique even self-dual Lorentzian lattice II$_{1,9}$ \cite{CS}.
The cosmological billiard describing the asymptotic behaviour near 
a spacelike singularity is based on the identification of the 
ten diagonal metric degrees of freedom of $D=11$ supergravity 
with the ten non-compact directions of a Cartan subalgebra (CSA) 
of $E_{10}$. This intriguing link between $D=11$ supergravity and 
$E_{10}$ was deepened in \cite{DaHeNi02,DaNi04} where it was shown 
that the bosonic equations of motion of $D=11$ supergravity at some
given spatial point, when restricted to zeroth
and first order spatial gradients in the metric and the three-form,
can be matched with the equations of motion of a one-dimensional
$E_{10}/K(E_{10})$ $\SI$-model restricted to levels $\ell\leq 3$.
In terms of the heights of the roots of $E_{10}$ involved in these 
correspondences, the Weyl-chamber billiard retains only roots of height
one ({\it i.e.} simple roots), while the work of \cite{DaHeNi02,DaNi04} 
has established the correspondence with (null) geodesic motion 
on the coset space $E_{10}/K(E_{10})$ up to height 29 included.
These results underline the potential importance of $E_{10}$, whose
appearance in the reduction of $D=11$ supergravity to one dimension
had been conjectured already long ago in \cite{Ju81,Ju82}, as a candidate
symmetry underlying M theory. A similar, but conceptually different,
proposal was made in \cite{West}, where $E_{11}$ (or some even larger 
symmetry containing $E_{11}$ \cite{West1}) has been suggested as a 
fundamental symmetry of M theory. Let us also note that links between 
the dynamics of gravitational theories and geodesic motions on certain
Lorentzian signature spaces were noticed long ago \cite{DeWitt:1967yk,GT}. 
For recent recent work in this vein in the context of homogeneous 
cosmologies (but not related to $E_{10}$), see \cite{TW,RT05}.

The substantial increase in the height of the roots entering
the correspondence between the two dynamics led to the key 
conjecture of \cite{DaHeNi02}, according to which the 
one-dimensional bosonic $E_{10}/K(E_{10})$ $\SI$-model is equivalent 
(or `dual') to the bosonic sector of $D=11$ supergravity, possibly 
augmented by further M-theoretic degrees of freedom. To establish 
this conjecture, one faces two challenges, namely to understand

\begin{itemize}
\item how the higher order spatial gradients are realized in the 
$\SI$-model, and how the full (untruncated) bosonic sector of $D=11$ 
supergravity is thereby embedded into the $E_{10}/K(E_{10})$ coset 
space dynamics; and
\item the meaning and significance of the imaginary (lightlike and timelike) 
roots of $E_{10}$ and their relation to M theory degrees of freedom
and M theory corrections of the `low energy' $D=11$ supergravity action.
\end{itemize}

Concerning the first challenge, it was shown in \cite{DaHeNi02} that
$E_{10}$ contains three infinite towers of Lie algebra elements, which 
possess the correct structure for representing the higher spatial gradients
of the metric, and of the `electric' and `magnetic' components of the
four-form field strength. However, it is still unknown how this 
identification could work in detail. The progress reported in the 
present paper concerns the second challenge. Namely, we shall show 
that certain (partially) known higher-order M theory corrections 
to the two-derivative, `effective' $D=11$ supergravity Lagrangian 
can be related, in a certain approximation which we shall explain, 
to special imaginary roots of $E_{10}$. Our `botanical' approach 
in searching for new connections between M theory and the hyperbolic 
Kac-Moody algebra $E_{10}$ is based on the working hypothesis that
{\em the $E_{10}/K(E_{10})$ $\SI$-model of \cite{DaHeNi02} not only describes 
M theory degrees of freedom beyond those of $D=11$ supergravity, but in 
addition contains hidden information about M theory corrections to the 
low energy effective theory (i.e. $D=11$ supergravity) at arbitrarily 
high orders.} More specifically, we shall consider the leading (8th order 
in derivatives) corrections to the usual supergravity Lagrangian and exhibit
their connection with certain imaginary roots of $E_{10}$. Different routes 
for interpreting the imaginary roots of $E_{10}$ in terms of brane
dynamics have been suggested in \cite{BrGaHe04}, and, in the framework 
of the $E_{11}$ proposal of \cite{West}, in \cite{Englert2003,EHL}.

In \cite {DaHeNi03} it was shown that the gravitational Hamiltonian,
or more precisely the Hamiltonian constraint at a given spatial point,
can be written in the form 
\beq \label{V4}
\HC(\B^a, \pi_{a},Q,P) = \tilde{N} \left[\frac12 G^{ab} \pi_a \pi_b  +
\sum_A c_A (Q,P,\partial\B,\partial Q)\exp\big(- 2 w_A (\B)\big)\right]
\eeq
with the rescaled lapse $\tilde{N} \equiv N/ \sqrt{g}$, where $g$ is 
the determinant of the {\em spatial} metric. Here $\B^a,\pi_a$ are the 
canonical variables corresponding to the diagonal (spatial) metric degrees 
of freedom, and $G^{ab}$ the (Lorentzian) `superspace' metric induced
by the Einstein-Hilbert action. $(Q,P)$ denote the remaining canonical
degrees of freedom associated to off-diagonal metric and various matter 
degrees of freedom, and $\partial\B, \partial Q$ their {\em spatial} 
gradients. The sum runs over various exponential potential `walls', which 
are indexed by $A$. The above form of the Hamiltonian is in particular 
valid for the bosonic sector of $D=11$ supergravity, in which case 
$a,b= 1,\dots, 10$. 

On the other hand, for the one-dimensional `geodesic'  $\SI$-model over 
the infinite dimensional coset space $E_{10}/K(E_{10})$, an analogous 
expression for the Hamiltonian constraint was also derived in \cite{DaHeNi03},
viz.
\beq\label{HKM}
H(\B^a,\pi_a, \N, p) =  n  \left[  \frac12  G^{ab} \pi_a \pi_b
      + \sum_{\A\in\Delta_+} \sum_{s=1}^{{\rm mult}(\A)}
      (\Pi_{\A ,s}(\N,p))^2 \exp\big(-2\A(\B)\big) \right]
\eeq
where $(\B^a,\pi_a)$ now denote the ten diagonal CSA degrees of freedom
of $E_{10}$, and $(\N, p)$ stand for infinitely many off-diagonal 
(Iwasawa-type) canonical variables, on which the quantities $\Pi_{\A,s}$ 
depend. The metric $G^{ab}$ now denotes
 the restriction of the unique invariant
bilinear form on $E_{10}$ to the CSA \cite{Kac}. This metric
happens to be {\em identical
with the metric appearing in} \Ref{V4} for the bosonic sector of $D=11$ 
supergravity; its explicit form is given in \Ref{pp} below. The sum on 
the r.h.s. of \Ref{HKM} ranges over all positive roots $\A$ of $E_{10}$
with their multiplicities [$={\rm mult}(\A)$], and the number $\A(\B)$ 
is the result of applying the linear form (=root) $\A$ to the CSA 
element $\B$. The formula \Ref{HKM} is actually valid for any $G/K(G)$ 
$\SI$-model over an indefinite Kac-Moody group $G$ with `maximal compact
subgroup' $K(G)$. Varying \Ref{HKM} w.r.t. the (new) lapse $n$ implies 
that the resulting solution is a {\em null geodesic} on this infinite 
dimensional coset manifold.

The similarity of \Ref{V4} and \Ref{HKM} is obvious, but the
coincidence of the two expressions and of the associated bosonic
equations of motion has so far only been established for a limited 
number of terms. The first check consists in matching those terms in 
\Ref{V4} and \Ref{HKM} which dominate the dynamics near a cosmological 
singularity. In this \textit{leading BKL approximation} one can prove 
\cite{DaHeNi03} that the `off-diagonal' degrees of freedom $(Q,P)$ 
in \Ref{V4} and $(\N,p)$ in \Ref{HKM} both `freeze' as one approaches 
the singular initial hypersurface $T=0$. [Here $T$ denotes the proper time. 
We shall reserve the letter $t$ to denote the coordinate time in the 
gauge $\tilde{N} = 1$, \textit{i.e.} $N= \sqrt{g}$. This coordinate 
time goes to $+\infty$ as $T \to 0$.] The precise form of the coefficient 
functions $c_A(Q,P,\partial\B,\partial Q)$, which is very complicated, 
does not matter in this BKL limit. What matters is that the coefficients of
the {\em leading contributions} are all \textit{non-negative}. That is,
we have $c_{A'}\geq 0$, where the primed index $A'$ labels the leading 
terms. This implies that, for large $\B$, the Hamiltonian \Ref{V4} takes 
the limiting form
\beq\label{V5}
{\HC}_{\infty}(\B^a, \pi_{a}) = \frac12 G^{ab} \pi_a \pi_b  +
\sum_{A'} c_{A'} \exp\big(- 2 w_{A'} (\B)  \big)
\eeq
where the sum is only over the `dominant walls'. It was also shown, that 
in this limit the exponential (Toda) walls appearing in (\ref{V5}) can 
be replaced by infinitely high (`sharp') walls. Technically, this means 
that one can replace the exponential functions in  (\ref{V5})
by an infinite step function $\Theta$ \cite{DaHeNi03}.
The billiard then takes place in the chamber defined 
by $\Theta(-2w_{A'}(\beta))=0$, or equivalently, $w_{A'}(\beta) \geq 0$.
An analogous argument applies to the coset dynamics \Ref{HKM}.
The freezing of all the off-diagonal degrees of freedom $\N,p$
implies that the combinations $\Pi_{\A ,s}(\N,p)$ also freeze near
the singularity, so that the Hamiltonian \Ref{HKM} takes the
limiting form (in the gauge $n=1$)
\beq\label{HKMlimit}
H_{\infty}(\B^a, \pi_{a}) = \frac12 G^{ab} \pi_a \pi_b  +
  \sum_{\A_i }  \Pi_{\A_i}^2  \exp \big(- 2 \A_i(\B)  \big)
\eeq
where the sum over the positive roots of $E_{10}$ can be asymptotically
restricted to the subset of leading positive roots $\A_i$.
It is easily seen that these `leading' positive roots are nothing
but the simple roots of $E_{10}$.

The equivalence between the two limiting dynamics \Ref{V5} and 
\Ref{HKMlimit} is now a consequence of the fact that the set of 
dominant supergravity walls $w_{A'} (\B)$ coincides with the set of
simple roots $\A_i(\B)$ of $E_{10}$ \cite{DaHe01}. The validity of 
this equivalence was extended in \cite{DaHeNi02,DaNi04} from the 
height one simple roots to height 29 in $E_{10}$ roots~\footnote{The
\textit{height} of a positive root $\A = \sum n_i \A_i$ ($n_i\geq 0$)
is defined as ht$(\A_i) = \sum n_i$ \cite{Humphreys,Kac}.
Correspondingly, the height of a negative root $\A = -\sum n_i \A_i$ ($n_i\geq 0$) is defined as $- \sum n_i$.}. At this
order, one keeps more exponential terms in both \Ref{V4} and \Ref{HKM}.
Now, the precise form of the supergravity coefficient functions 
$c_A(Q,P,\dots)$ {\em does} matter, and was shown in \cite{DaHeNi02,DaNi04} 
to correspond precisely to the form of the coset functions 
$\Pi_{\A ,s}(\N,p)$, with a `dictionary' relating the variables of
both dynamics. This matching between the two Hamiltonians was 
checked at the level of the equations of motion.
It is still unclear whether (and how) one can extend the dictionary
relating the dynamical variables in both models so as to prove 
the equivalence between the two Hamiltonians \Ref{V4} and \Ref{HKM} 
beyond height 29. It is quite possible that this dictionary becomes 
spatially non-local in the gravity variables, and that the simple 
identification assumed above between the diagonal gravitational
degrees of freedom $\B^a , \pi_{a}$
and the $E_{10}$ CSA variables entering \Ref{HKM} must be modified
beyond height 29. Such non-trivial changes of variables may be needed, 
for instance, in order to reconcile the fact that the only negative 
contribution in \Ref{HKM} comes from the CSA (the first term on 
the r.h.s.), whereas in the expansion of the gravity Hamiltonian 
\Ref{V4} there may arise negative contributions also from (subleading) 
terms with $c_A <0$ --- as follows from considering, for instance, 
homogeneous spatial geometries with curvature of arbitrary sign.

Our aim here is to investigate what types of wall forms can be 
formally associated with higher order terms (linked to quantum 
corrections in M-theory), notably the leading ones of the type 
$R^4, R^2 (DF)^2$, $R (DF)^3$ and $(DF)^4$, as well as the topological 
Chern Simons term $A_3 \wedge R \wedge R \wedge R \wedge R$.
Remarkably, we shall find that these terms correspond, in leading
approximation, to \textit{time-like} imaginary roots of $E_{10}$.
Recall that the `length' of a root is calculated by means of
a quadratic form $G^{a b}$, which is the same as the one appearing in
\Ref{HKM} (and \Ref{V4}), and which, for indefinite Kac-Moody
algebras, is {\em Lorentzian}. Spacelike roots, for which 
$ G^{a b} \A_a \A_b \equiv \A^2 > 0$, are called `real roots'
in the mathematical literature, while light-like or time-like ones, 
for which $\A^2 \leq 0$, are called `imaginary roots' \cite{Kac}.
All the roots up to height 29, that have been explicitly checked to 
match between the two Hamiltonians \Ref{V4} and \Ref{HKM} turned out to 
be {\em real} and {\em positive}. By contrast, we will find here that
the leading roots associated with 8th-order derivative corrections 
are, at once, {\em imaginary} ($\A^2 \leq 0$)
and {\em negative} ( {\it i.e.} of the form
$\A = -\sum n_i \A_i$ with $n_i\geq 0$). For the leading $R^4$ corrections,
they lie deep inside the root-space lightcone, with $\A^2= -10$, 
at height ht$(\A) = -115$, and at level $\ell = -10$.

The appearance of exponential walls $\exp \big(- 2w(\B)\big)$ associated 
with \textit{negative roots} in a version of the gravitational Hamiltonian 
\Ref{V4} with higher order corrections included, might seem to be
in plain contradiction with the structure \Ref{HKM} of the coset 
Hamiltonian, which contains a sum restricted to \textit{positive roots}
$\A\in\Delta_+$. However, one must remember that the explicit form of the 
coset Hamiltonian strongly depends on the parametrisation used for a 
general coset element $ \VC \in E_{10}/K(E_{10})$. The form \Ref{HKM}
results from an Iwasawa parametrization $ \VC = \DC \BC$, where 
$\DC$ is `diagonal' (\textit{i.e.} in the Cartan torus of $E_{10}$)
and $\BC$ of Borel, or  `upper triangular' type, \textit{i.e.} obtained 
by exponentiating a sum of positive root generators. If instead
we were to employ a more complicated parametrization of a generic
coset element $ \VC \in E_{10}/K(E_{10})$, namely involving,
besides the exponentiation of infinitely many positive root
generators, the exponentiation of some\footnote{We note that such
`mixed' non-triangular parametrizations of infinite dimensional 
Kac-Moody coset spaces have so far not been considered in
the literature. It seems likely that, for the exponentiation 
to be well defined for hyperbolic Kac-Moody groups, one can only 
have a sparse number of negative root generators mixed with an 
infinite number of positive root ones.} negative root generators,
one would end up with an Hamiltonian of a more general form than \Ref{HKM}. 
The latter would involve some mixture between positive-root walls
$\exp \big(- 2 \A_A(\B)  \big)$ (with $\A_A\in\Delta_+$), and negative-root 
ones $\exp \big(+ 2\A_B(\B)\big)$ (with $\A_B\in\Delta_+$). Actually, 
the mixtures of positive-root walls and negative-root ones arising
in such non-Borel parametrizations are rather complicated, and do not 
reduce to a decoupled sum of normal exponentials $\exp \big(-2\A_A(\B)\big)$, 
and inverse ones $\exp \big(+2\A_B(\B)\big)$. A non-Borel parametrization 
for the root $\A_B$ tends to generate superpositions of effects proportional 
to both $\exp \big(+ 2\A_B(\B)\big)$ and $\exp \big(- 2 \A_B(\B)\big)$. 
Therefore, the wrong-sign walls $\sim\exp\big(+ 2 \A_B(\B)\big)$ can be 
expected to dominate over the usual-sign ones only in some special 
corner of the dynamical space. Such an approximate dominance is enough 
for our purposes here, because we shall see below that the
negative root walls associated to $R^4$ terms can only be meaningfully 
considered in some intermediate domain of the dynamical evolution.

No doubt there is still a long way towards the ultimate goal of 
proving the equivalence between the (essentially unique) $E_{10}$ coset 
dynamics \Ref{HKM} and the full effective (bosonic) supergravity action 
(involving the quantum effect of further M-theoretic degrees of freedom), 
but we hope that the results reported here will allow one to make new 
and stringent tests of our basic conjecture, and to make new 
predictions concerning both the structure of higher-order corrections 
in supergravity and M-theory, which are completely inaccessible by 
conventional methods, and the existence of special $\gl$-representations 
in the level decomposition of $E_{10}$.

\section{Asymptotics of frames and scale factors}

When working with $D=11$ supergravity fields, we use Lorentz 
(tangent space) tensors throughout, with  the `mostly plus' metric 
$\eta^{AB} = (-+ \cdots +)$, and the following index conventions
\bea
\mbox{Flat spacetime indices} \quad &:& \quad A,B,C,... \in\{0,1,\dots,10\} \nn
\mbox{Flat spatial indices} \quad &:& \quad a,b,c,... \in \{ 1,\dots,10\}
\eea
When needed, the coordinate spacetime indices will be denoted as
$M,N,P,Q$ and the coordinate space ones as $m,n,p,q$.

The Lorentz covariant derivative is defined in the usual manner on
any Lorentz vector $V_A$ via
\beq
D_A V_B := \pa_A V_B + {\OM_{AB}}^C V_C
\eeq
where $\pa_A$ is a (non-commuting) frame derivative linked to 
the usual (commuting) spacetime derivative $\pa_M$ through 
$\pa_A \equiv {E_A}^M \pa_M$, where ${E_A}^M$ is the elfbein.
The spin connection is given by the standard formula
\beq
\OM_{A\,BC} = \frac12 \Big( \OO_{ABC} - \OO_{BCA} + \OO_{CAB} \Big)
            = - \OM_{A\ CB}
\eeq
in terms of the coefficients of anholonomicity
\beq
{\OO_{AB}}^C := {E_A}^M {E_B}^N \big( \pa_M {E_N}^C -\pa_N {E_M}^C \big)
              = - {\OO_{BA}}^C
\eeq
where ${E_M}^A$ is the co-frame, \textit{i.e.} the inverse of the 
elfbein ${E_A}^M$. Expressed in flat indices, the Riemann tensor is
\bea
R_{ABCD} &=& \partial_A \OM_{B\, CD} - \partial_B \OM_{A\, CD} 
   + {\OO_{AB}}^E \OM_{E\, CD} \nn
&& + {\OM_{A\, C}}^E \OM_{B ED} - {\OM_{B\, C}}^E \OM_{A ED}
\eea
Likewise, for the 4-form field strength we will also use flat indices 
such that $F_{ABCD} = {E_A}^M {E_B}^N {E_C}^P {E_D}^Q F_{MNPQ}$ , etc.

For the further analysis, we will rely on the following
(zero-shift) 1+10 split of the elfbein
\beq
E_M{}^A =
\left(\begin{array}{ccc}
          N &  0\\
          0 & e_m{}^a
\end{array}\right)
\eeq
with the spatial (inverse) zehnbein ${e_m}^a$, and adopt the gauge 
\beq\label{n}
N= \sqrt{g} \equiv {\rm det} \, {e_m}^a 
\eeq
As explained in \cite{DaHeNi03}, and as is also obvious from the 
definition of $\tilde{N}$ in \Ref{V4}, one consequence of this 
choice is that the leading kinetic term in \Ref{V4} simplifies 
to $\propto G^{ab} \pi_a \pi_b$. Splitting indices into $A=0$ and 
$A=a$, where 0 denotes a `flat' (proper) time index and $a$ a flat 
spatial index, the coefficients of anholonomicity become
\bea\label{OO}
\OO_{abc} &=& 2{e_{[a}}^m {e_{b]}}^n \partial_m e_{nc} \quad , \nn
\OO_{0bc} &=& N^{-1} {e_b}^n \partial_t e_{nc} \quad , \nn
\OO_{a00} &=& \OM_{0\, 0a} = - {e_a}^m N^{-1} \partial_m N
   = - {e_a}^m e^{-1} \partial_m e
\eea
with all other coefficients of anholonomicity vanishing. From the above
formulas we obtain in particular
\beq
\OM_{0\, bc} =  N^{-1} {e_{[b}}^n \partial_t e_{nc]} \quad ,\qquad
\OM_{a\, b0} =  N^{-1} {e_{(a}}^n \partial_t e_{nb)}
\eeq
where $(...)$ and $[...]$ denote symmmetrization and antisymmetrization
with strength one, respectively.

In the sequel we will analyze the leading asymptotic behaviour of 
various fields near $T=0$ as functions of the logarithmic scale 
factors $\B$. To this aim we introduce, as an intermediate object, 
the Iwasawa co-frame ${\theta_{m}}^a \equiv \NC^a_m$, where $\NC^a_m$ 
is the upper triangular matrix entering the Iwasawa decomposition of the 
metric $g_{m n}$ that was introduced in \cite{DaHeNi03}. [This triangular 
matrix has ones on the diagonal, so $\det\theta =1$.]  In terms of 
this intermediate frame, the ten-dimensional spatial Cartan moving 
frame (zehnbein) is given by (no summation on $a$) 
\beq\label{frame}
 {e_{m}}^a = e^{-\B^a} {\theta_{m}}^a
\eeq
in each spatial section $t= const.$ (recall that we use a zero-shift gauge). 
The line element then reads
\beq
ds^2 = - (N dt)^2 + \sum_{a} (e^{-\B^a} \theta^a)^2
\eeq
The decomposition \Ref{frame} of the zehnbein is most useful in analyzing
the small $T$ behavior of the gravity model. More specifically, the 
dynamical variables $\B^a$ and ${\theta_{m}}^a$ behave very differently
as $T \to 0$. While the frame ${\theta_{m}}^a$ was shown to `freeze' 
(\textit{i.e.} to have a well-defined limit) as $T\rightarrow 0$, the 
logarithmic scale factors $\B^a$ (whose time-derivatives are proportional 
to the Kasner exponents) generically diverge toward $+ \infty$ near the
singular initial (`big bang') spatial hypersurface $T=0$, while exhibiting
chaotic oscillations of BKL type. [Note, in passing, that we never assume
the $\B$'s to be only functions of time $t$. All our supergravity dynamical 
variables are functions of all eleven coordinates $(t,x^m)$]. With these 
notations, the gauge condition \Ref{n} implies
\beq
N  = \exp \left(-\sum_{a=1}^{10} \B^a\right) \det\theta
  = e^{-\SI}
\eeq
while the $D=11$ volume density behaves as
\beq\label{E}
E = N\sqrt{g} = e^{-2\SI}
\eeq
Here $\SI$ denotes the sum of all logarithmic scale factors,
\beq
\SI:= \sum_{a=1}^{10} \B^a
\eeq
This combination will play an important role in the remainder. As $ T \to 0$, 
the volume factor $\sqrt{g} = e^{-\SI}$ tends toward zero, and $\SI$ tends 
to plus infinity. Similarly, one can exhibit the behavior of various 
other fields near the singular initial hypersurface by factoring them
into a part that `freezes' as $T\rightarrow 0$ and is independent of the
scale factors, and another, dependent on the $\B$'s, which diverges
and may oscillate near $T=0$. 

Making use of \Ref{frame} we can compute the asymptotic behavior of
the coefficients of anholonomicity and the spin connection. Suppressing 
inessential prefactors, we find for the connection components with 
one (proper) time index
\bea\label{oab0}
\OM_{0\, bc} &\sim& 0 \nn
\OM_{a\, b0} &=&  - N^{-1} \D_{ab} \pa_t \B^a + \dots =
 - \D_{ab} e^{+\SI} \pa_t \B^a + \dots
\eea
where the symbol $\sim$ indicates that we keep only the leading
contributions (from \Ref{OO} it is easily seen that that the components 
with {\em two} time indices are sub-dominant). For the purely spatial
components, we get for the anholonomicity coefficients\footnote{The
spatial indices in these and other formulas below always refer to
the basic orthonormal frame ${e_m}^a$, whereas the barred quantities
associated to the intermediate (non-orthonormal) frame ${\theta_m}^a$
are only used as auxiliary objects.}
\beq\label{Oabc}
\OO_{abc} = e^{\B^a + \B^b - \B^c} \bar\OO_{abc} +
\D_{ac} e^{\B^b} \bar\pa_b \B^c - \D_{bc} e^{\B^a} \bar\pa_a \B^c
\eeq
where $\bar\pa_a \equiv {\theta_a}^m \pa_m$, and
\beq
\bar\OO_{abc}:= 2{\theta_{[a}}^m {\theta_{b]}}^n \partial_m \theta_{nc}
\eeq
Hence the spatial components of the spin connection are given by a
sum of terms:
\bea
\OM_{abc} &=& \frac12 \Big( e^{\B^a + \B^b - \B^c} \bar\OO_{abc} -
e^{\B^b + \B^c - \B^a} \bar\OO_{bca}  + e^{\B^c + \B^a - \B^b}
\bar\OO_{cab}\Big) + \nn && \qquad
\; +\;\D_{ac} e^{\B^b} \bar\pa_b \B^c - \D_{ab} e^{\B^a} \bar\pa_c \B^c
\eea 
It is important that the barred quantities $\bar\OO_{abc}$ and the frame
derivatives $\bar\pa_a$ all possess finite limits as $T\rightarrow 0$; they
correspond to the `frozen' components in the language of \cite{DaHeNi02}.
Hence, for $T\sim 0$
\beq\label{oabc}
\OM_{abc} \propto e^{\B^a + \B^b - \B^c} + 
\mbox{permutations of the indices $a,b,c$}
\eeq
where we may have $a=b$ or $a=c$ (but $b \neq c$ as $\OM_{a \, bc}
= - \OM_{a \, cb}$). We will make use of these formulas in section~5,
when determining the asymptotic behavior of the curvature and other
`composite' quantities of interest.

\section{Dominant walls and $E_{10}$ roots}

In this section, we recall some basic facts and useful formulas
concerning the Lie algebra $E_{10}$ and its roots. A crucial fact
here is the identification of the logarithmic scale factors appearing 
in \Ref{frame} with the CSA degrees of freedom of $E_{10}$. This will
enable us to re-interpret the exponential walls $\exp[-2w(\B)]$ in 
the gravitational Hamiltonian in terms of the exponential (Toda) 
walls $\exp[-2\A(\B)]$ entering the $E_{10}$ $\SI$-model \Ref{HKM}.
As mentioned already, the scalar product on the space of scale factors 
induced by the Einstein-Hilbert action (for $D=11$) coincides with the 
restriction of the invariant bilinear form on $E_{10}$ to its CSA 
\cite{DaHeNi03}. This scalar product is given by
\beq\label{Gab}
(\B|\B) \equiv G_{ab} \B^a \B^b = \sum_{a=1}^{10} \big( \B^a \big)^2
  - \left( \sum_{a=1}^{10} \B^a \right)^2
\eeq
It is indefinite (Lorentzian) due to the lower unboundedness of the
Einstein-Hilbert action under variations of the conformal factor.
As generally shown in \cite{DaHeNi03}, in the limit $T\rightarrow 0$ 
the dynamics of the scale factors is well approximated by a billiard. 
The region defining this billiard is contained in the forward lightcone 
in $\B$-space, and bounded by `sharp walls'; the latter are hyperplanes 
defined as the zeros of certain linear forms (`wall forms') 
\beq\label{w}
w(\B) = \sum_{a=1}^{10} p_a \B^a
\eeq
The metric on the dual space of wall forms is the contravariant form
of \Ref{Gab}
\beq\label{pp}
(p|p) \equiv G^{ab} p_a p_b = \sum_{a=1}^{10} p_a^2  
  - \frac19 \left( \sum_{a=1}^{10} p_a \right)^2
\eeq
and this is also the metric appearing in \Ref{V4} and \Ref{HKM}.
Below, we will also use the notation
\beq\label{p}
w \equiv (p_1, \dots ,p_{10})
\eeq
to designate a particular wall form \Ref{w}. In \cite{DaHeNi03} we have 
explained how to determine these wall forms from the matter coupled 
Einstein-Hilbert action in a canonical formulation (see also the following 
section). A main result was that for $D=11$ supergravity, each wall
form $w$ can be associated with a particular real root of $E_{10}$, in
accordance with the fact that the root space is dual to the CSA, which itself
is identified with the space of scale factors $\B$. In particular,
the {\it dominant} walls are associated with the {\it simple roots}. 
In the above basis \Ref{p}, the simple roots of $E_{10}$ are
\bea\label{simpleroots}
\A_0 &=& (1,1,1,0,0,0,0,0,0,0) \nn
\A_1 &=& (-1,1,0,0,0,0,0,0,0,0) \nn
\A_2 &=& (0,-1,1,0,0,0,0,0,0,0) \nn
\A_3 &=& (0,0,-1,1,0,0,0,0,0,0) \nn
\A_4 &=& (0,0,0,-1,1,0,0,0,0,0) \nn
\A_5 &=& (0,0,0,0,-1,1,0,0,0,0) \nn
\A_6 &=& (0,0,0,0,0,-1,1,0,0,0) \nn
\A_7 &=& (0,0,0,0,0,0,-1,1,0,0) \nn
\A_8 &=& (0,0,0,0,0,0,0,-1,1,0) \nn
\A_9 &=& (0,0,0,0,0,0,0,0,-1,1)
\eea
and one easily checks that they generate the $E_{10}$ Dynkin diagram
via the scalar product \Ref{pp}. In the billiard picture, the dominant
wall forms are therefore 
\bea\label{simplewalls}
w_{\A_0} (\B) \equiv \A_0 (\B) &=& \B^1 + \B^2 + \B^3 \nn
w_{\A_1} (\B) \equiv \A_1(\B) &=& \B^{2} - \B^1 \nn
  &:& \nn &:& \nn
w_{\A_9} (\B) \equiv \A_9(\B) &=& \B^{10} - \B^9 
\eea
The first simple root corresponds to a leading `electric wall', while the
remaining nine simple roots correspond to the leading `symmetry walls'.
All other walls (electric and symmetry, as well as the magnetic and 
gravitational walls) are subdominant and can be written as positive
linear combinations of these leading walls. Consequently, the billiard 
takes place in the fundamental Weyl chamber, namely the `wedge' defined 
by the conditions $\A_i (\B)\geq 0$, or
\beq
\B^1 + \B^2 + \B^3 \geq 0 \quad , \qquad
\B^1 \leq \B^2 \leq \dots \leq \B^{10}
\eeq
Every positive root of $E_{10}$ can be expressed as
a linear combination of simple roots:
\beq\label{root}
\A= \ell\A_0 + \sum_{j=1}^9 m^j \A_j \equiv 
[l\, ; \, m^1 , m^2 , m^3 , m^4 , m^5 , m^6 , m^7 , m^8 , m^9]
\eeq
where, of course, $\ell, m^j \geq 0$, and the height of $\A$ is
${\rm ht}\, (\A) = \ell + \sum_j m^j$. The brackets ($[\cdots]$) 
denote the components of a root in the basis of simple roots, and
are to be distinguished from the use of round parentheses used to 
denote the components of a root in the `coordinate basis' of the $\B$'s.
The integer $\ell$ is referred to as the `$A_9$ level', or simply the 
`level' of the root $\A$. (Different `levels' can be defined w.r.t. other 
subalgebras of $E_{10}$ such as $D_9$ \cite{KleNic04} or $E_9$ \cite{KMW}).
This level provides a grading of the Kac-Moody algebra $E_{10}$,
allowing a decomposition of the $E_{10}$ Lie algebra into an infinite 
tower of $\gl$ representations, which has been worked out up to 
level $\ell \leq 28$ in \cite{NiFi03}. 
The two bases \Ref{p} and \Ref{root} are related by
\bea
p_1 &=& \ell - m^1 \nn
p_2 &=& \ell + m^1 - m^2 \nn
p_3 &=& \ell + m^2 - m^3 \nn
p_4 &=& m^3 - m^4 \; , \; \dots \; , \;
p_9 = m^8 - m^9 \quad , \quad p_{10} = m^9
\eea

Roots corresponding to wall forms associated with particular higher 
order terms will appear at a fixed level $\ell$. For the associated 
roots we will usually give the one corresponding to the root of lowest 
height within a given $\gl$ multiplet, for which the integers $p_a$ are 
ordered according to
\beq
p_1 \geq p_2 \geq \cdots \geq p_{10} 
\eeq
Because this is also the lowest weight of the corresponding $SL(10)$
representation, we can read off the associated Dynkin labels directly
from the tables of \cite{NiFi03}, as well as the outer multiplicity $\mu$,
{\it i.e.} the number of times this representation occurs in the
decomposition of $E_{10}$. Starting from the lowest weight, other
weights in the representation can be reached by successive addition of the 
`symmetry roots' $\A_1, \dots , \A_9$. From \Ref{simpleroots} it is evident
that this operation simply amounts to a permutation of the $p_i$'s
(and in in terms of the scale factors $\B^a$ to a replacement of the 
spatial indices $a$ by other spatial indices). The highest weight, 
which corresponds to a root of maximal height, is reached when the 
order above has been inverted to $p_1 \leq \cdots \leq p_{10}$. 

As an example let us briefly recall how the gravitational billiard 
walls arise by expanding out the Hamiltonian constraint following from 
the Einstein-Hilbert action, as described in detail in \cite{DaHeNi03},
and how they can be identified with particular $E_{10}$ roots. Using
\Ref{Oabc}, they are obtained by considering the following contribution 
(no summation on repeated indices)
\beq
E \Omega_{abc} \Omega_{abc} = e^{-2\SI + 2\B^a + 2\B^b - 2\B^c} 
\bar\OO_{abc}\bar\OO_{abc} \quad + \;\; \quad \mbox{subleading terms}
\eeq
Because the walls always appear via exponentials $e^{-2w(\B)}$,
the relevant linear form is 
\beq\label{wgrav}
w^{grav}_{abc} = \SI -\B^a -\B^b +\B^c,  \qquad \mbox{with $a \neq b$}
\eeq 
Choosing the indices $(a,b,c)= (9,10,1)$ so as to obtain the root 
of lowest height in this multiplet [with Dynkin label (100000010)]
we get
\beq\label{wgrav'}
w^{grav}_{9\, 10\, 1} =  (2,1,1,1,1,1,1,1,0,0) \quad\Longleftrightarrow\quad
\A = [3\, ;\, 1 , 3 , 5 , 4 , 3 , 2 , 1 , 0 , 0]
\eeq
In particular the associated root is {\em real} ($\A^2 = +2$) and 
{\em positive} in our conventions. Consequently, for $D=11$ supergravity,
the above gravitational walls arise at level $\ell=3$ and are thus `behind'
the level $\ell =1$ electric walls defined by the simple root $\A_0$.
This contrasts with pure gravity, where the leading gravitational wall 
always corresponds to a simple root of the algebra $AE_n$ \cite{DaHeJuNi01}.

\section{Higher order corrections: what is known}

Still as a preparation, we here summarize briefly what is known about the 
8th order derivative corrections to the eleven-dimensional supergravity 
action, which are of the type $R^4$, $R^2 (DF)^2$, $R (DF)^3$, $(DF)^4$, 
$\cdots, F^8$, where $R$ stands for the curvature (Riemann) tensor, and 
$F = dA$ is the four-form field strength (there are no $R^3 DF$ terms because 
any possible contraction of the indices will involve an $\E$-tensor, and 
vanish by the Bianchi identity for either $R$ or $DF$). In addition, 
there are separate 8th-order derivative terms of Chern Simons type
proportional to $A_3 R^4$ required for the cancellation of anomalies
\cite{DuffLiuMinasian}. The supersymmetric completion of these 8th-order 
terms was studied in \cite{PeetersVanhove} and, in a superspace formulation, 
in \cite{Cederwall,HoweTsimpis} (see also the review \cite{Howe}).
The presence of such corrections can be inferred $(i)$ from (string) 
one-loop corrections to $D=10$ type IIA string amplitudes 
\cite{GreenSchwarz,SakaiTanii}, which are similar in form to $(\A')^3$ 
corrections at string tree level 
\cite{GrossWitten,GrisaruZanon,Myers,GroSlo87}; 
$(ii)$ from considering M-theory corrections at one loop 
\cite{GreenVanhove1,GreenVanhove2, GGK,Tseytlin}, $(iii)$ from
the computation of 4-graviton amplitudes in supermembrane theory 
by use of supermembrane vertices \cite{DHN,PNPW,Plefka}; or $(iv)$ from 
the structure of divergences in $D=11$ supergravity \cite{DS}. However, 
none of these approaches gives the complete `reduced' ({\it i.e.}
maximally simplified) kinematical structure of all the (bosonic) 
8th-order derivative corrections. The paper \cite{DS} does contain 
explicit expressions for all terms $\propto R^M (DF)^N$ ($M+N=4$),
but in an `unreduced' form involving the Bel-Robinson tensor, which 
for our present purposes is rather cumbersome~\footnote{Beware of a 
missing factor 12 in the third term, $\propto P^2$, on the r.h.s. 
of Eq. (5a) of the first reference in \cite{DS} (J.~Plefka, private 
communication).}. Moreover, none of these papers gives expressions
for the terms involving the undifferentiated 4-form $F$, such as $F^8$.

We note that the $D=11$ results should be compatible with those of
the IIA theory, whenever the $D=10$ results can be `lifted' to $D=11$.
In this case one must worry about the possible existence of terms 
that algebraically vanish in $D=10$, but not in $D=11$. As the number 
of independent invariants made out of the Weyl tensor is $= 7$ when 
$D \geq 8$ \cite{Fulling}, this is not a concern for the curvature terms. 
However, the situation is more subtle for invariants built from the 
covariant derivative $DF$ of the 4-form field strength, whose number 
does depend on the dimension. For instance, the number of $(DF)^4$ 
invariants is found to be $= 23$ for $D=9$, and $= 24$ for $D=11$, 
as can be conveniently checked by use of the computer algebra 
package LiE \cite{LiE}. For $D=10$, we would have 29 invariants
with a 4-form field strength; however, there are only 15 invariants
that can be built from the NSNS 3-form field in IIA supergravity 
(which means that we cannot simply lift the results of \cite{GroSlo87} 
to eleven dimensions).

Yet another subtlety one must keep in mind is that the lightcone results 
do not necessarily agree with results of a `covariant' calculation. 
A case in point is the `Euler-Lovelock' term which we will encounter 
below: this term does not show up in the lightcone calculation of the 
4-graviton amplitude, hence is not determined, whereas it does appear 
in the covariant approach of \cite{Tseytlin,PeetersVanhove}. However, 
this term should start to contribute to on-shell amplitudes from 5-point 
amplitudes onwards, whence it should become visible in a lightcone 
computation of the 5-graviton amplitude.

We are here only interested in {\em on-shell counterterms}. That is, we 
shall ignore all terms proportional to the equations of motion because 
they can be absorbed into redefinitions of the basic fields. For this
reason, the Riemann tensor $R_{ABCD}$ can be replaced by the Weyl
tensor $C_{ABCD}$, since all terms containing the Ricci tensor or 
Ricci scalar can be absorbed (modulo $F$-dependent terms) into a 
redefinition of the metric by use of the equations of motion. For the 
$DF$ type terms, we can similarly discard the trace and the fully 
antisymmetric combinations because
\beq
 D^A F_{ABCD} \sim 0 \quad , \qquad D_{[A} F_{BCDE]} = 0
\eeq
by the equations of motion and the Bianchi identity (we neglect
terms $\propto F^2$ on the r.h.s. of the equation of motion;
we shall see later that they are indeed subdominant).
Thus we must only deal with the irreducible representations
of the rotation group represented by the following Young tableaux:
\beq
C_{ABCD}  ~\; \sim \;~ 
\young(\hfil\hfil,\hfil\hfil)  
\quad , \qquad
D_A F_{BCDE}  ~\; \sim \;~ 
\young(\hfil\hfil,\hfil,\hfil,\hfil)
\eeq

We first focus on the terms quartic in $R$ (or $C$), because, 
as we shall see, they are the dominant ones within our analysis. From 
Refs. \cite{Tseytlin, PeetersVanhove} these terms are given by the 
expression (modulo a \textit{positive} coefficient)
\beq\label{R4tot}
\LC^{(4) {\rm tot}} = E ( J_0 - 2 \, \IC_2)
\eeq
where $E = \det {E_M}^A = N \sqrt{g}$, 
\beq \label{j0}
J_0 := t_8 t_8 R^4 + \frac14 E_8 \equiv X - \frac18 Z
\eeq
and
\beq \label{i2}
\IC_{2} = 
  \frac14 E_8 + 2 \E_{11} A_3 \left[ \tr R^4 - \frac14 (\tr R^2)^2\right]
\equiv - \frac18 Z + 2 \E_{11} A_3 \left[ \tr R^4 - \frac14 (\tr R^2)^2\right]
\eeq
Here, we use the same condensed, index-free notation as in \cite{Tseytlin}, 
while the notation $X\equiv t_8 t_8 R^4$ and $Z\equiv - 2 E_8$ is used 
in \cite{PeetersVanhove}. \footnote{Note a misprint in the definition 
of Z given in the appendix B of \cite{PeetersVanhove}: in Eq.(B.14) 
$\E_{10} \E_{10}$ should be replaced by $- \E_{10} \E_{10}$. The sign 
which appears in the {\tt hep-th} version of that paper is correct.
We are grateful to P.~Vanhove for a discussion on this point.}

In particular, $t_8$ denotes the 8-index tensor that enters the 
kinematic factor of both tree-level and one-loop 4-particle 
string amplitudes \cite{GreenSchwarz,Schwarz82}. The normalization  
of $t_8$ is such that the contraction of $t_8$ with four times the 
same antisymmetric matrix $M$ yields
\beq \label{t8}
t_8 M M M M := 24 \, \tr M^4 - 6 \, (\tr M^2)^2
\eeq
$E_8 \equiv \frac1{3!}\E_{11} \E_{11} R R R R$ is shorthand for the 
so-called `Euler-Lovelock density'; more explicitly, writing out all 
indices for once, we have
\bea \label{e8'}
E_8 &=& \frac1{3!}\, \E^{ABCD_1\dots D_8} \E_{ABCE_1\dots E_8} 
   {R_{D_1 D_2}}^{E_1E_2} \cdots  {R_{D_7D_8}}^{E_7 E_8} \nn
    &=& -  \, \delta^{D_1\dots D_8}_{E_1\dots E_8}
   {R_{D_1 D_2}}^{E_1E_2} \cdots  {R_{D_7D_8}}^{E_7 E_8}
\eea
where $\delta_8^8 = +1, -1, 0$ is (when non-zero) the signature of
the permutation between $D_1\dots D_8$ and $E_1\dots E_8$.
Observe that in Minkowskian signature the product $\E_{11} \E_{11}$
contains a \textit{minus} sign with respect to the usual light-cone
term entering string-theory amplitudes $\E_{8} \E_{8} = + \delta_8^8$.

There are many subtle issues related to the relative sign between 
the $t_8 t_8 R^4 = X$ and the $\frac14 E_8 = - \frac18 Z$ contributions. 
In particular the usual lightcone calculation of 4-graviton amplitudes 
\cite{GreenSchwarz} yields the \textit{same} combination $X-Z/8$ 
for the IIA theory at tree level and at the one-string-loop level. 
However, according to Refs. \cite{Tseytlin, PeetersVanhove}, the 
correct one-string-loop term (which lifts to the M-theory result) 
differs from the light cone calculation by an Euler-Lovelock term
(and the Chern Simons term), and should read $X- Z/8 + 2 (Z/8) = X + Z/8$
instead (this is precisely the opposite sign from the one obtained 
in the lightcone computation of \cite{GreenSchwarz}). To recover 
(or check) this result in the lightcone gauge one would have to 
calculate a 5-graviton amplitudes in IIA string theory at one loop, 
but we are not aware of such computations. This issue is important 
because, as we shall see below, our cosmological billiards are 
\textit{a priori} sensitive to the Euler-Lovelock term, and hence to 
the flip of sign between the lightcone result $X -Z/8$, and the 
complete result $X- Z/8 + 2 (Z/8) = X + Z/8$. 

In the following, we shall decompose the total contribution \Ref{R4tot} 
into three separate terms: the term $L^{(4) 1} = J_0
= t_8 t_8 R^4 + \frac14 E_8$, the Euler-Lovelock
term $L^{(4) 2} = - \frac12 E_8 =+ 2 \cdot \frac18 Z$ contained in 
$- 2 \, \IC_2$, and the Chern-Simons term $L^{CS}$ contained in $- 2 \, \IC_2$.
The first term \Ref{j0} (which coincides with the lightcone result $X -Z/8$) 
can be worked out more explicitly \cite{Tseytlin,PeetersVanhove, PNPW}, 
modulo terms that either vanish on-shell, or are equivalent to sub-leading 
terms (\textit{e.g.} $\sim R^3 F^2$)
\bea \label{C4}
\LC^{(4) 1} &=& E J_0 = E \left(X - \frac18 Z\right) \nonumber \\
&=& 192 E \Big(- C^{ABCD} {C_{AB}}^{EF} {C_{CE}}^{GH} C_{DFGH} \nonumber\\
&&\qquad\qquad \; + \; 4 C^{ABCD} {{{C_A}^E}_C}^F {{{C_E}^G}_B}^H C_{FGDH}\Big)
\eea
The second term $\LC^{(4) 2}$ has a much more complicated form when
expressed in this way \cite{PeetersVanhove,DS}, and we will therefore
leave it `in the $\E$-form' \Ref{e8'}.
 We shall discuss the effects of this term
in more detail in section~7, and of the Chern Simons term in section~8.

The mixed terms containing $(DF)$ factors have so far not been 
reduced to such a simple form. Let us just note that, in the
(eleven-dimensional) light-cone
gauge, the determination of {\em all} these terms can be reduced 
to an $\E$ trace over $SO(9)$ $\GA$-matrices, and, more specifically, 
can be obtained by contracting any four factors from
\beq
C_{\A\B\C\D} := \GA^{ij}_{\A\B} \GA^{kl}_{\C\D} C_{ijkl} \quad ,\qquad
(DF)_{\A\B\C\D} := \GA^{ij}_{\A\B} \GA^{klm}_{\C\D} D_i F_{jklm}  
\eeq
with $\E^{\A_1 \dots \A_{16}}$, as follows by inspection of the relevant
vertex operators in \cite{GGK,DHN}. Here the indices $i,j,\dots = 1,...,9$ 
label the nine {\em transverse} dimensions in eleven dimensions,
and $\GA^{ij}$ and $\GA^{ijk}$ are the standard $SO(9)$ $\GA$-matrices 
with spinor indices $\A,\B = 1,\dots , 16$. However, we would expect the 
{\em caveats} concerning the difference between light-cone and `covariant' 
amplitudes, mentioned above for the $C^4$ contributions, 
also to apply to the mixed terms.

\section{Scaling behaviour of curvatures and field strengths}

Next we exhibit the leading asymptotic behaviour of the various fields 
and their `composites' corresponding to the higher order corrections
near $T=0$, as functions of the logarithmic scale factors $\B$,
making use of the decomposition \Ref{frame} and other formulas 
derived in section~2. Exploiting the identification of the scale factors 
with the CSA degrees of freedom, we shall then try to interpret the 
exponential walls $\exp[-2w(\B)]$, as obtained from the higher order  
corrections, in terms of new exponential (Toda) walls $\exp[-2\A(\B)]$ in 
the $E_{10}$ $\SI$-model. To this aim, let us first determine the leading 
asymptotic behaviour of the curvature, starting with that of the spin 
connection $\OM$ and the anholonomicity coefficients $\OO$. Because the 
Iwasawa frame freezes in this limit, the `leading order' amounts to 
neglecting the time derivatives of the intermediate frame, \textit{i.e.} 
terms proportional to $\pa_t \theta^a$. More precisely, the terms 
$\pa_t \theta^a$ are proportional to some exponential walls, starting 
with some simple root walls $\sim \exp( - 2 \A_i(\B))$. A consequence 
of this fact is that when discussing the subleading terms in the
8-th order correction (which are smaller than the dominant
term by a factor $\sim \exp\Big( - 2 \sum n_i \A_i(\B)\Big)$ 
with $n_i\geq 0$), one should, at some stage, take into account
their `mixing' with the terms proportional to $\pa_t \theta^a$
and coming from the leading curvature contribution.

To proceed we now substitute the results derived in section~2 into the 
expressions for the Riemann tensor in order to see in more detail how 
its various components behave in the limit $T\rightarrow 0$. Retaining 
only the relevant terms containing time derivatives, we get
\bea
R_{0a0b} &=& - \pa_0 \omega_{ab0} - \Omega_{0ac} \omega_{cb0} + \dots = \nn
&=&  \D_{ab} e^{2\SI} 
\big[ \pa_t^2 \B^a + \pa_t\SI \pa_t \B^a - (\pa_t \B^a)^2 \big] + \dots
\eea
for the components with two 0's, and
\beq\label{R0abc}
R_{0abc} \propto N^{-1} e^{\B^a + \B^b - \B^c} \equiv 
e^{\SI + \B^a + \B^b - \B^c} \quad + \; \mbox{ permutations of $a,b,c$}
\eeq
for the component with one 0. For the purely spatial components we have
\beq\label{Rabcd}
R_{abcd} = R^{(1)}_{abcd} + R^{(2)}_{abcd}
\eeq
where the first term on the r.h.s. contains the contributions with
$\omega_{ab0}$
\beq\label{Rabcd1}
R^{(1)}_{abcd} := \omega_{ac0}\omega_{bd0} - \omega_{ad0} \omega_{bc0}
= 2\D_{c[a} \D_{b]d} e^{2\SI} \pa_t \B^a \pa_t \B^b
\eeq
while the second term $R^{(2)}_{abcd}$ involves all contributions from the
purely spatial spin connection \Ref{oabc} and is proportional to
\bea\label{Rabcd2}
R^{(2)}_{abcd} &\propto& e^{\B^a + \B^b + \B^c - \B^d} \quad\qquad
  \, \mbox{or}\nn
&\propto& e^{\B^a + \B^b + \B^c + \B^d - 2\B^e} \quad
  \; \mbox{or}\nn
&\propto& e^{\B^a + \B^b + 2 \B^e - \B^c - \B^d} 
\eea
together with all permutations of the indices $a,b,c,d$; note that 
some of these indices may coincide. The index $e$ is independent of
$a,b,c,d$; it is the summation index in the terms quadratic in the 
spin connection. For later reference, we note that the combinations 
appearing in the exponentials on the r.h.s. of \Ref{Rabcd2} differ 
from one another by linear combinations of `symmetry roots' $(\B^a - \B^b)$, 
corresponding to the so-called `symmetry walls' \cite{DaHeNi03}.

As a check of our computation of the leading curvature components, 
let us work out the Einstein-Hilbert action to leading order; we get
\bea\label{EH}
E R &\simeq& e^{-2\SI}\left(\sum_{a,b} R^{(1)}_{abab} - 
 2\sum_a R_{0a0a}\right) \nn
  &=& - 2\pa_t^2 \SI + \sum_a (\pa_t\B^a)^2 - (\pa_t \SI)^2
  \equiv - 2\pa_t\SI + \sum_{a,b} G_{ab} \pa_t \B^a \pa_t \B^b
\eea
Therefore, modulo a total time derivative $- 2\pa_t^2 \SI$, we recover 
the kinetic terms for the diagonal metric components in \Ref{V4}, 
with the usual (Lorentzian) `superspace' metric $G_{ab}$ from \Ref{Gab}.
It is the inverse ($=G^{ab}$) of this metric which appears in both the 
Hamiltonians \Ref{V4} and \Ref{HKM} in the Introduction, and was written 
out in \Ref{pp}.

It is easily seen that the leading components of the curvature are 
$R_{a0a0}$ and $R^{(1)}_{abab}$ (with $a \neq b$, and no summation on
repeated indices). These leading components are of order $e^{2\SI}$ 
and can be written as (remembering that, in our gauge, $\B^a$ is 
approximately piecewise linear in the cordinate time $t$)
\bea \label{leadingR}
R_{a0a0} &=& e^{2\SI} v_a \bar{v}_a \nn
R^{(1)}_{abab} &=& e^{2\SI} v_a v_b  \qquad {\rm for} \, a \neq b
\eea
where, for shortness of notation, we have defined
\beq
v_a := \pa_t \B^a \quad , \qquad 
\bar{v}_a := \sum_{b\neq a} \B^b \equiv \pa_t \SI -\pa_t \B^a
\eeq
For the Kasner solution, the coordinate time can be taken to be
$t = -\ln T$, and the `velocities' $v_a$ are then just the 
Kasner exponents.

The remaining components of the curvature tensor are sub-leading 
in the sense that they are suppressed in comparison with the leading 
term (of order $\sim e^{2\SI}$) by exponentially small factors. 
Before discussing in a unified way how one can define a relative 
ordering between several contributions, let us collect various 
results for the {\em ratios} between subleading curvature (and 
$F$-dependent) terms, and the leading curvature $R \sim e^{2\SI}$.
From \Ref{R0abc} and \Ref{Rabcd2} we see that
\beq\label{R0}
R_{0abc} \DD e^{2\SI} \sim e^{-\SI + \B^a + \B^b - \B^c}
+ \ {\rm permutations}
\eeq
Notice that the combination appearing in the exponential on the
r.h.s. is just (minus) the gravitational wall form \Ref{wgrav}. Similar 
estimates hold for $R^{(2)}_{abcd}$; a representative example is
\beq\label{R2}
R^{(2)}_{abcd} \DD e^{2\SI} \sim e^{-2\SI + \B^a + \B^b + \B^c - \B^d} 
\eeq
and all other such quotients differing only by symmetry walls.

The asymptotic behaviour of the 4-form and of its derivatives can be 
analyzed in a similar manner. For the electric field strength we have
\beq
F_{0abc} = e^{\SI - \B^a - \B^b - \B^c} \bar F_{tabc}
\eeq
where 
\beq
\bar F_{tabc} := \theta_{ma} \theta_{nb} \theta_{pc} {F_t}^{mnp}
\eeq
and the overall factor $e^\SI$ comes from the inverse lapse $N^{-1}$.
[We recall that the indices $0,a,b,c$ denote orthonormal-frame indices.]
As shown in \cite{DaHeNi03}, it is precisely the barred field strength
$\bar F_{tabc}$ (in the gauge $N=\sqrt{g}$; or, in any gauge, the
conjugate momenta $N \sqrt{g}\,  F^{tmnp}$) which possesses a limit as 
$T\rightarrow 0$. Comparing  the electric field strength
with the square root of the leading curvature
$ R^{1/2} \sim e^{\SI}$, we obtain [no summation on $(a,b,c)$]
\beq\label{Fe}
F_{0abc} \DD e^{\SI} \sim e^{-(\B^a + \B^b + \B^c)}
\eeq
where the combination $(\B^a + \B^b + \B^c)$ appearing in the exponential 
on the r.h.s. is related to an `electric wall', see \cite{DaHeNi03}.
In particular, for $(a,b,c)=(1,2,3)$ we just re-obtain the simple root
$\A_0$ in \Ref{simpleroots}, which explains the term `electric root'.
We thus conclude that the electric field strength is `smaller' than the 
square root of the dominant curvature by one electric wall.

For the magnetic field strength, we have similarly
\beq
F_{abcd} = e^{\B^a + \B^b + \B^c + \B^d} \bar F_{abcd}
\eeq
with
\beq
\bar F_{abcd} :=
  {\theta_a}^m {\theta_b}^n {\theta_c}^p {\theta_d}^q  F_{mnpq}
\eeq
Again the barred field strength is the one that remains finite on
the initial hypersurface. Note the different positions of the world
indices in these two definitions. [The correct index position 
for the quantities that freeze is easy to remember as the quantities 
that freeze are the $Q$'s and the $P$'s: $A_{mnp}$ and their spatial 
derivatives, and their conjugate momenta $\pi^{mnp}$.] Comparing the 
 magnetic field strength to the square root of the leading curvature,
  we obtain
\beq\label{Fm}
F_{abcd} \DD e^{\SI} \sim e^{-\sum_{e\neq a,b,c,d} \B^e}
\eeq
The combination appearing in the exponential on the r.h.s. is now
a magnetic wall, which is also equal to the sum of {\em two} 
(non-overlapping) electric walls. Consequently, the magnetic field
strength is `smaller' than the electric field strength by one 
electric wall, and `smaller' than the square root of the leading 
curvature term by two electric walls.
This shows that all the higher order terms which are algebraic in 
$R$ and $F$ are sub-dominant compared to the leading $R^4$ terms. 
For instance, $F^8$ is `smaller' than the leading $R^4$ term by
eight electric walls.

Finally, the covariant derivative of the field strength is dominated 
by the time derivative of an electric component:
\beq
D_0 F_{0abc} = e^{2\SI - \B^a -\B^b - \B^c} D_t \bar F_{tabc} + \dots
\eeq
Hence,
\beq\label{DF}
D_0 F_{0abc} \DD e^{2\SI} \sim e^{- (\B^a + \B^b + \B^c)}
\eeq
In other words, $DF$ is smaller than $R$ by one electric wall.

We can conveniently summarize the relative ordering of the
building blocks entering the 8th order correction by introducing
the symbol $\preceq$ to denote the ordering of
exponential factors involving simple wall forms. As will be
further discussed at the beginning of Section 6, we can
treat the negative exponentials of the simple roots
as small parameters: $\E_i \sim \exp( - \A_i(\B)) \preceq 1$.
This defines an ordering such that $n_i \geq n'_i$ implies
$ \exp( - \sum n_i \A_i(\B)) \preceq \exp( - \sum n'_i \A_i(\B))$.
Now, by looking more closely at the various exponential factors
appearing on the r.h.s's of the comparison equations
\Ref{R0}, \Ref{R2}, \Ref{Fe}, \Ref{Fm}, \Ref{DF} above,
one sees that they can be rewritten as negative exponentials
of sums of simple roots. For instance,
the r.h.s. of \Ref{R0} is $e^{-(\SI - \B^a - \B^b + \B^c)}$,
where the combination $\SI - \B^a - \B^b + \B^c$ is the
{\em gravitational} wall form, see \Ref{wgrav} above,
which can be written as a sum of simple roots, see Eq. \Ref{wgrav'}.
Actually, it is convenient, for a quick perusal of the
relative ordering of various contributions, to neglect
the ordering with respect to symmetry roots $\A_1, \dots, \A_9$,
see \Ref{simplewalls} in the next Section, and to
focus on the ordering defined by the number $n_0$ of
leading electric roots $\A_0(\B) = \B^1 +\B^2 +\B^3 $.
In other words, we shall simply write
$ \exp( - \sum n_i \A_i(\B)) \preceq \exp( -  n_0 \A_0(\B))$,
and use the ordering
$ \exp( -  n_0 \A_0(\B))\preceq \exp( -  n'_0 \A_0(\B))$
when $ n'_0 \geq n_0$. For instance, in view of
the fact that the gravitational wall form which
appears in \Ref{R0} is at level $\ell =3$ (see \Ref{wgrav'}),
we can write $e^{-(\SI - \B^a - \B^b + \B^c)} \preceq e^{-3\A_0}$. 
Finally, with this notation, the above results simplify to
\beq \label{R0'}
R_{0abc}\DD e^{2\SI} \preceq e^{-3\A_0}
\eeq
\beq \label{R2'}
R^{(2)}_{abcd} \DD e^{2\SI} \preceq e^{-6 \A_0}
\eeq
\beq
F_{0abc}^2 \DD e^{2\SI} \preceq e^{-2\A_0} \quad , \qquad
F_{abcd}^2 \DD e^{2\SI} \preceq e^{-4\A_0}
\eeq
(in comparing with the curvature which is second order in derivatives,
we must square the 4-form field strengths because they have only one
derivative)
\beq\label{DF1}
DF \DD e^{2\SI} \preceq e^{-\A_0}
\eeq

These estimates show that the dominant terms among the various 8th-order
corrections $R^4$, $R^2 (DF)^2$, $R (DF)^3$, $(DF)^4$, $\cdots, F^8$
are the curvature terms $R^4$. Among the $R^4$ terms,
the leading contributions will come either from the time-time components 
$R_{0a0a}$, or from the purely spatial ones $\sim R^{(1)}_{abab}$. Then, 
the next to leading contributions will come either from $R^2 (DF)^2$ 
or from $R^3 F^2$, and will be smaller than the leading one by a 
factor $\sim e^{-2\A_0}$.

\section{Leading behavior at higher orders: curvature terms}

Before proceeding with the analysis of the higher order derivative 
contributions in the Lie algebraic setting, let us clarify in which 
regime our analysis of their effects, and in particular of the 
8-th order derivative corrections, will be meaningful. This regime  
is one of an `intermediate asymptotics' of the type 
$ T_P \ll T \ll T_0$. Here, $T_P$ is the (eleven-dimensional) 
Planck time, and $T_0$ is some initial (or final) time away from 
the big crunch (or big bang) singularity located at $T=0$. We assume 
that the Cauchy data at $T_0$ are such that spatial derivatives of 
the metric and the 3-form are mildly smaller than their time 
derivatives, say $ \pa_a g < 0.2  \pa_T g$. We also assume that 
the time derivatives correspond to curvatures much smaller
than the eleven-dimensional Planck curvature $T_P^{-2}$.
Then, for a while, the BKL-type analysis becomes better and better
for times $T \ll T_0$, in the sense that the potential walls
become sharper and sharper, and that the Weyl chamber of $E_{10}$ 
appears as a dominant billiard description of the dynamics of the 
diagonal components of the metric at any given spatial point. 
Such a billiard description of the Einstein-three-form system 
remains accurate until the higher-order corrections to the
action, $\sim R + T_P^6 R^4 + \cdots$ become important. This
happens for curvatures $ R \sim T_P^{-2} $, \textit{i.e.}
for cosmological times $ T \sim T_P$. As there are no small
coupling parameters in M-theory, when the leading correction
$\sim R^4$ becomes important, all higher-order ones $\sim
T_P^{2(N-1)} R^N$ become also important as $T\sim T_P$. 
However, in the intermediate regime $ T_P \ll T \ll T_0$, 
it is meaningful to consider only the terms $\sim R^4$
and to treat them as a small correction to the usual supergravity
billiard defined by $S_0 = R + F^2 + A F F$. In other words,
in such a regime we can meaningfully combine two separate expansion 
schemes: $(i)$ a {\em height expansion} in which the negative 
exponentials of the simple roots are treated as small parameters,
say $\E_i \sim \exp( - \A_i(\B))$, and $(ii)$ a {\em small curvature
expansion} in which $ \E_c \sim T_P^2 R \sim (T_P/T)^2$ is treated
as a separate small parameter. Note that the basic rule
behind the first expansion is to order the exponential terms
by their height. Indeed, a generic wall $\exp ( - 2 \A(\B))$,
with a positive root $\A = \sum n_i \A_i$ is seen as the
product of small numbers, $ \prod (\E_i)^{2 n_i}$, which gets 
smaller and smaller as the $n_i$ increase. Another, and more
heuristic, way to think about the height expansion would be to
replace the ten individual $\B^a$'s, or better their simple-root 
combinations $\A_i(\B)$, by some `average scale factor' $\oB$, or 
some `average simple root' $\A_s \equiv \overline{\A}_i$. This is
a legitimate approximation because the $\A_i(\B)$ increase toward
$+ \infty$ and remain positive during their chaotic motion within 
the Weyl chamber $\A_i(\B) \geq 0$. In terms of such averaged
quantities one can estimate the effect of a generic wall 
$ \exp( -2 \sum n_i \A_i (\B))$ as being on average equivalent to  
$ \exp(-2 (\sum n_i) \A_s )= \exp(- 2{\rm ht} (\A)\,\A_s )$.

As we have seen in Section  4, the dominant terms, in the sense 
of the height expansion just explained, among the 8th order derivative 
corrections $R^4$, $R^2 (DF)^2$, $R (DF)^3$, $(DF)^4$, $\cdots, F^8$
are the curvature terms $R^4$. Therefore, these are the terms 
that will define the leading corrections to the dynamics in the 
intermediate asymptotics $ T_P \ll T \ll T_0$.

Let us examine more carefully the contribution of the terms $R^4$
in the Hamiltonian constraint $\HC$. In any gauge an additional
term $J$ in the (invariant) Lagrangian density contributes the
term $ - E J$ in the Hamiltonian constraint. We recall from \Ref{E}
that the eleven-dimensional volume density behaves as
$E = N\sqrt{g} = e^{- 2 \SI}$. As we have seen above, the 
leading frame components of the curvature blow up
proportionally to $e^{2\SI}$. Hence, a term $J \sim R^4$
contributes $ - E J \sim e^{- 2 \SI} (e^{2\SI})^4 \sim e^{6\SI}$.
[Certain special quartic curvature invariants $R^4$ might
only contribute  terms $ \ll (R_{0a0a})^4$, but we shall see 
below that both the $t_8 t_8 R^4$ and $E_8$ do contain terms 
$\sim(e^{2\SI})^4 $.] For more generality, let us consider 
a correction containing an arbitrary power of the curvature, say 
$J \sim R^N$ (the precise kinematical structure of $R^N$ does 
not matter at this point). This contributes to the Hamiltonian 
constraint a term
\beq \label{RN}
E R^N \propto e^{2(N-1)\SI}
\eeq
The important point for our analysis is that, in the Hamiltonian 
constraint \Ref{V4} such a term has the form of an exponential 
wall $ \sim e^{-2w(\B)}$, where $w(\B)$ is a linear form in the $\B$'s.
The  associated `wall form' $w_N \equiv w[R^N]$ is easily read off, viz.
\beq \label{wRN}
 w_N (\B) = - (N-1) \SI \quad\Longleftrightarrow\quad
 w_N = - (N-1)\cdot \big(1,1,1,1,1,1,1,1,1,1\big)
\eeq
Let us now see for which values of $N$ the wall form \Ref{wRN}
happens to be a root of $E_{10}$. A simple \textit{necessary}
criterion for checking whether a wall form $w$ might be a root of 
$E_{10}$ consists in computing the squared length of $w$ with the 
scalar product of the CSA of $E_{10}$ \Ref{pp}. Indeed, all forms 
$w$ on the CSA which belong to the $E_{10}$ root lattice\footnote{We 
recall that the {\em root lattice}, \textit{i.e.} the set of integer 
combinations of simple roots, contains, but is strictly larger than, 
the set of roots.}, and which are such that $w^2 = - 2 j$, for some
integer $j \geq 0$,  are (imaginary) roots of $E_{10}$ and 
{\it vice versa}. For the wall form \Ref{wRN}, one easily finds
\beq
w_N^2 =  - \frac{10}9 (N-1)^2
\eeq
This is negative, and is an even integer only when $N=3k+1$,
with some positive integer $k$. This includes the case of
main interest here, namely $N=4$, \textit{i.e.} $k=1$.

It is easily checked that when the necessary condition $N=3k+1$
is met, the corresponding $ w_{3k+1} = - 3 k \SI$ does belong to the 
root lattice, and is therefore an (imaginary) root. For instance, 
the quartic correction $R^4$ corresponds to the $E_{10}$ root
\bea\label{rootN}
 w_4  &=& - (3,3,3,3,3,3,3,3,3,3) \;\; \Longleftrightarrow \;\; \nn
\A &=& - [10\, ;  7 , 14 , 21 , 18 , 15 , 12 , 9 , 6 , 3]
\eea
which is of squared length $\A^2 = -10$  and height ht$(\A)= - 115$;
its level is $\ell= - 10$. We see that this root lies rather deep 
inside the lightcone in root space. More generally, for 
$N=3k+1$ (\textit{i.e.} for corrections $R^4, R^7, R^{10}, \cdots$), 
one gets wall forms corresponding to $E_{10}$ roots of squared length 
$-10 k^2$, height ${\rm ht}(\A)= -115k$, and level $\ell = -10k$
for $k=1,2,3,\dots$.

Looking back at the potential wall \Ref{wRN} associated to the 
correction $R^N$ we notice something special. Barring special
cancellations, all terms in the correction $R^N$ lead to a unique
dominant potential wall $e^{2(N-1)\SI}$. This is quite different 
from what happens for the usual walls that have appeared in the
analysis of \cite{DaHeNi02,DaHeNi03}. Indeed, the walls associated 
to the electric energy of the 3-form, or its magnetic energy, or to 
the `gravitational energy' all came in multiplets of the group of permutations 
of spatial indices. For instance, the electric walls were of type 
$\exp( - 2 (\B^a + \B^b +\B^c))$, where $a,b,c$ are three different 
spatial indices. The lowest-height electric root $\A_0 =\B^1 + \B^2 +\B^3$
is accompanied by other electric roots at the same level $\ell =1$, 
which are obtained by acting on $\A_0$ by the generators of $SL(10)$ 
(with roots $\B^2 - \B^1, \cdots$), which have the effect of
permuting the $SL(10)$ indices among $ a=1,2, \cdots, 10$.
This fact was associated to the fact that the basic electric root 
$\A_0 =\B^1 + \B^2 +\B^3$ was simply the lowest weight associated 
to a generator of $E_{10}$ which was given by an antisymmetric 3-tensor
$SL(10)$ representation, say $E^{abc}$. In other words, the walls
which come in permutation multiplets are associated to
non-trivial tensor representations of $SL(10)$.

The `isotropy' of the wall form \Ref{wRN}, \textit{i.e.} the fact
that it does not depend on the choice of a special permutation of 
$SL(10)$ indices, suggests that $w_N$, Eq. \Ref{wRN}, should be 
associated to a root of $E_{10}$ whose associated generator is a
$SL(10)$ {\em singlet}. We do find in the tabulated results of 
\cite{NiFi03} that there are indeed the requisite singlets matching 
the $ R^4$ terms at level $\ell =10$, and the predicted $R^7$ terms
at level $\ell =20$. Moreover, these are the {\em only singlets} among the
$4\,400\,752\,653$ representations identified up to level $\ell \leq 28$
in \cite{NiFi03}. In view of the `contravariant' character of the 
`raising generators' of $E_{10}$, namely, $ E^{(3)} \equiv E^{abc}$, 
$E^{(6)} \equiv E^{a_1 \dots a_6} \sim [E^{(3)},E^{(3)}]$, {\it etc.} 
it is easy to see that $SL(10)$  singlets can only occur at levels 
$\ell = 10 k$ (so that one can use $3 k$ covariant  $\E$ tensors
of $SL(10)$ to soak up the $3 \ell$ contravariant indices of
$E^{( 3 \ell)}$) \footnote{We thank Axel Kleinschmidt
for pointing out this algebraic fact.}. It is, however,
remarkable that: $(i)$ singlets {\em do occur} at these
levels with non zero outer multiplicities, and $(ii)$
there exists such a nice compatibility between the
algebraic structure of $E_{10}$ (built by `commuting'
three-form level-1 generators $E^{abc}$), and the wall structure 
entailed by curvature corrections. See the final Section
for further discussion of this point.

Moreover, among the highest weights of the representations at levels 
$\ell=10$ and $\ell = 20$ the associated root \Ref{rootN} and its integer 
multiples are the ones of largest root height. {\em Therefore, for any $N$, 
among the roots associated with the $2N$-th order derivative corrections, 
\Ref{rootN} is always the root that lies deepest inside the past lightcone.} 
For the moment, no data are available beyond $\ell > 28$, but we 
confidently predict the requisite $SL(10)$ singlets to appear at 
levels $10k$ with roots $3 k \SI$ and non-vanishing outer multiplicities.

We can summarize our findings in a little lemma.

\vspace*{0.2cm}\noindent
{\bf{Lemma:}} {\it Corrections proportional to the $N$-th power of the 
Riemann (or Weyl) tensor are compatible with $E_{10}$ if and only
if $N=3k+1$, {\it i.e.} for $R^4, R^7, R^{10}, \dots$. Moreover, such
corrections must be associated with $\gl$-singlet generators of $E_{10}$ 
at levels $\ell = - 10 k$.}

\vspace*{0.2cm}

For the special value $k=0$, the `leading term' is just the kinetic
term \Ref{EH}, and there is no `leading wall'. For $k=1,2$, {\it i.e.}
$N=4,7$, the outer multiplicities of the singlets are known: $\mu = 3$ 
for $\ell =10$, and $\mu = 913$ for $\ell =20$ \cite{NiFi03}. We have no 
physical interpretation for these values in terms of the $ R^4$ and $R^7$ 
corrections to the action. It might have been tempting to associate 
the outer multiplicity $\mu = 3$ to the three different 
contributions ($ t_8 t_8 R^4, E_8$ and the Chern-Simons term) entering 
\Ref{R4tot}. However, we shall see that the Chern-Simons one is 
subdominant. Note also that for $E_{11}$ the root corresponding to
\Ref{rootN} is associated with the $SL(11)$ representation $(0000000003)$, 
which is no longer a singlet.

Below, we will show that this result extends to corrections of mixed 
type $R^{N-n} (DF)^n$. Remarkably, the very same constraint $N=3k+1$ 
on the allowed powers of curvature, which we found here assuming 
compatibility with the $E_{10}$ root lattice, was arrived at by 
a very different argument in \cite{Russo} (to wit, an expansion in 
integer powers of the inverse membrane tension). Note also that, as far 
as we know, no detailed information is yet available about the
$R^N$ corrections for $N>4$.

\section{Positivity properties of leading curvature terms}

Let us comment on two interesting aspects of the $E_{10}$ root
\Ref{rootN}  associated with the leading $R^4$ corrections.
First, as announced in the Introduction the root \Ref{rootN},
besides being imaginary (or more precisely, time-like, $\A^2 < 0$), 
is also negative, $ \A \in \Delta_-$. Physically, this corresponds 
to the fact that the $R^4$ corrections grow faster than the leading 
Einstein terms near the singularity. As we already explained, this 
fact limits the validity of our analysis to some intermediate 
asymptotic $ T_P \ll T \ll T_0$. Within such an intermediate asymptotic 
expansion, we see no {\it a priori} conflict between the occurrence 
of negative $E_{10}$ roots in the M-theory action, and the form 
\Ref{HKM} of the coset action, involving only \textit{positive} 
roots of $E_{10}$. Indeed, as already pointed out above, the 
positive root form \Ref{HKM} is the result of using a special nilpotent 
(\textit{i.e.} purely upper triangular) parametrization of the coset 
$E_{10}/K(E_{10})$. The use of a more complicated parametrization, 
involving a mixture of upper and lower triangular parametrization, 
would entail the presence of negative roots in the coset action.

However, no matter how one parametrizes the solution, the trajectory
that extremizes the coset action is simply a null geodesic on
$E_{10}/K(E_{10})$, and therefore geometrically the same solution.
In the case of finite-dimensional
Lie algebras one can prove that a generic geodesic curve is conjugate
to one in the CSA. However, no proof of this result is known for
hyperbolic Kac-Moody algebras, nor is it known
whether such a statement holds at all, see \cite{Kac2}. If that were
true for geodesics on $E_{10}/K(E_{10})$ one would immediately
conclude that a generic null geodesic is conjugate to a null geodesic
in the CSA, {\em i.e.} to the curve
$ \B_a (t) = v_a t + \B_a^{(0)}$ (with $v^2 =0$) which
runs straight from `past null infinity' towards `future null infinity'
in the CSA. Fortunately, we can qualitatively
control the global behaviour of null geodesics without having to
assume the unproven equivalence with a CSA geodesic.\footnote{
See, in this respect,
 the Appendix below where we emphasize that null geodesics
on $E_{10}/K(E_{10})$ are `strongly chaotic' in the
sense of being extremely sensitive to small changes in their initial
conditions. This result suggests that geodesics which do not lie
within the CSA might be somewhat different from the simpler (Kasner)
CSA ones.} Indeed, we can remark that, in the Borel parametrization
of Eq. \Ref{HKM}, the potential terms $\propto \Pi^2$ are all
non-negative, so that they define a non-negative `squared-mass term',
$ -\pi^2 = M^2 \geq 0$ for the motion projected in $\B$-space.
As $\dot \B^a \propto \pi^a$ this
implies that the $\B^a$ worldline is everywhere timelike or null.
Remembering that, in terms
of the logarithmic scale factors $\B_a$'s, `future infinity' is
$ \SI \equiv \sum \B_a \to + \infty$, and corresponds to a `big crunch'
singularity where the volume $ \sqrt{g} = e^{-\SI} \to 0$, we conclude
that a generic null geodesic, which is initially future-directed, will never
be able to turn back to become past-directed.
In other words, this proves
 that a generic $E_{10}$ null geodesic cannot correspond to
a `bouncing' cosmological solution, where the volume of space reaches
a minimum before it re-expands. In view of this
argument, and if there is any truth to the $E_{10}$ conjecture,
we expect that the \textit{sign} of the higher-order corrections
to the supergravity action will be such as to allow
\textit{no bouncing solutions}.

We have studied the sign and magnitude of the various leading 
quartic corrections in \Ref{R4tot} to see if they might be
compatible with our expectations.\footnote{Strictly speaking, 
the sign of the leading correction is not logically related to 
the bounce/non-bounce behaviour. Indeed, even if the leading correction 
indicates a bounce behaviour, the next order corrections may 
prevent the completion of this bounce.} Actually, the analysis of 
the sign of the quartic-in-$R$ contribution to the gravity action
is rather involved, because the precise kinematical structure of
the $R^4$ now does matter. The first important fact is that the
sign of the coefficient of the loop correction \Ref{R4tot}
is {\em positive} \footnote{Due to a remnant 
Euclidean-type convention, one should reverse the overall
sign of the actions in \cite{Tseytlin}.}. Then there arises the issue
of the actual sign of the two leading contributions to the action,
\Ref{j0} and  the $E_8$ term in \Ref{i2} (which, as we shall 
see below, is dominant compared to the Chern-Simons one)
when they are evaluated `on-shell', \textit{i.e.} along a solution 
of the leading Einstein-3-form dynamics. If we can prove that the 
sum of the on-shell actions \Ref{j0} and  the $E_8$ term in \Ref{i2}
is numerically \textit{positive}, this will prove that
the corresponding contribution to the correction of
the gravity Hamiltonian is\footnote{We use here the usual
first order result $H_1 = - L_1$ for the correction to
the Hamiltonian corresponding to a correction to the
Lagrangian. This result applies to the case where one
introduces no new degrees of freedom, but only modifies
the action for given degrees of freedom, possibly
by higher derivative terms.} \textit{negative}. In that case, 
the wall associated to the $R^4$ corrections (characterized  by 
a negative, time-like root, with negative coefficient) will represent 
no longer a `mountain', which would obstruct the motion of $\B$ towards 
infinity and thereby induce a bounce, but instead a `canyon' or 
a `crevice' precipitating the collapse of space toward zero 
volume (\textit{i.e.} $\SI \to +\infty$).

Let us therefore have a closer look at the various $R^4$ contributions 
to the action.  We consider first the contributions coming from \Ref{j0},
which explicitly read as \Ref{C4}, and will later consider the $E_8$ 
one contained in \Ref{i2}.
The first observation is that in leading order (in the sense explained
above), only terms with $R_{a0a0}$ and $R_{abab}^{(1)}$ need to be
considered in \Ref{C4}. Secondly, it is straightforward to verify 
that all terms proportional to either $(C_{0a0a})^4$ or $(C_{abab})^4$ 
(no summation on indices!) cancel between the two contributions 
in \Ref{C4}. Substitution of \Ref{leadingR} into \Ref{C4} and some
further calculation then reveals for the leading part of the density 
\Ref{j0} the explicit expression
\beq \label{j0va}
E J_0 = 768 e^{6 \SI} V_2^4,
\eeq
where
\beq \label{j0vb}
V_2^4 \equiv  {\sum_{a,b}}' \,v_a^2 v_b^2 (v_a \bar{v}_a
+ v_b \bar{v}_b -\bar{v}_a\bar{v}_b)^2 +
\frac13 {\sum_{a,b,c}}'' \, v_a^2 v_b^2v_c^2 (v_a+ v_b +v_c)^2 \geq 0 
\eeq
and the sum $\sum'$ over $ a,b$ is restricted to $ a \neq b$, while
the sum $\sum''$ ranges over the three indices $a,b,c$ with the restriction
that $ a \neq b$, $ b \neq c$, and $ c \neq a$. Note that each
ordered pair $a< b$ contributes two equivalent terms to the sum $\sum'$,
while each ordered triplet $a<b<c$ contributes six equivalent terms
to the sum $\sum''$.
The explicit
expression \Ref{j0vb} as a sum of squares proves that $J_0 \geq 0$. 
If that were the only contribution to the action, this argument would 
suffice to prove the negativity of the corresponding term in the 
Hamiltonian. However, we still need to consider the term proportional 
to the Euler-Lovelock density in \Ref{i2}. Using the explicit 
expression of the Euler-Lovelock densities in Kasner spacetimes 
\cite{Deruelle},\footnote{After discussion with the author, it 
appears that the overall coefficient $2^{2 p -1}$
in Eqs. (22a) and (22b) of \cite{Deruelle} should be replaced
by the, {\it alas}, larger coefficient $ (2 p)!/(2 p -1)$.}
we get the following explicit, leading expression for
the relevant term in \Ref{i2}

\bea \label{e8}
\LC^{(4) 2} = -  \frac12 E  E_8 = +  \frac14 E Z
= - 46080 \left[ \pa_t (e^{6 \SI} V_7)
+ e^{6 \SI} V_8 \right]
\eea
where
\beq
V_7 := \sum_{a_1<\cdots <a_7} v_{a_1} \cdots v_{a_7}
\eeq
 and
\beq
V_8 := \sum_{a_1<\cdots < a_8} v_{a_1} \cdots v_{a_8}
\eeq
Note that, contrary to what happened for the sums $\sum'$
and $\sum''$ above, the sums appearing in $V_7$ and $V_8$
contain only one contribution per ordered multiplet of indices.
In the following, we shall discard the total time 
derivative\footnote{Note, however, in passing that if one
were to evaluate $\LC^{(4) 2}$ \textit{on-shell} before
discarding this term, it would yield a non-zero contribution
(but a contribution which is equivalent to a reparametrization
of the basic degrees of freedom).}  appearing as first term on the 
right-hand side of \Ref{e8}, and evaluate
\beq \label{e8'}
\LC^{(4) 2'} = - 46080 e^{6 \SI} V_8
\eeq

In order to see whether the term \Ref{e8'} is likely to modify 
the positivity of \Ref{j0vb} we need to compare $(46080/768) V_8 = 60 V_8$
to the term $V_2^4 $, \textit{i.e.} to the expression defined by 
the restricted sums $\sum', \sum''$ in \Ref{j0vb}. The latter 
quantity, $V_2^4 $, was shown above to be a sum of squares, 
and therefore to be always $\geq 0$. We need to
study the sign and magnitude of the additional contribution
$ - 60 V_8$.

We can obtain rather simply an approximate
knowledge of the magnitude of $V_8$ in the following way.\footnote{We
thank Ofer Gabber for suggesting this method for bracketing the
value of $V_8$.} We consider the 10-th degree polynomial
$P(x) = \prod_a (x - v_a)$. Let us consider on-shell values of the
$v_a$'s, i.e. such that $v^2 = - 2 \sum_{a<b} v_a v_b = 0$.
Let us also normalize the values of the $v$'s by imposing
$ \sum_a v_a = 1$. [So that we have both $\sum_a v_a = 1 =\sum_a v_a^2$].
In the expansion of the polynomial $P(x) = x^{10} - c_9 x^9 + c_8 x^8
+ \cdots + c_2 x^2 - c_1 x + c_0$, we then know $c_9 =\sum_a v_a = 1$,
$c_8 =\sum_{a<b} v_a v_b = 0$ and we have $c_2 = V_8$, where $V_8$
is the quantity we wish to bracket. Because \textit{all} the roots
of $P(x)$ are real, it is well known that \textit{all} the roots
of its successive derivatives $P'(x), P''(x), \cdots$ will also be real.
By considering the expansion in powers of $x$ of the second derivative
$P''(x)/90 =  x^8 - 0.8  x^7 + 0 x^6 + \cdots + V_8/45$, and
relating it to its real roots, say $x_i$ (for $ i = 1, \cdots, 8$),
we get the following expression for $V_8$:
$ V_8 = 45 \prod_i x_i$ where the $x_i$'s are constrained by
$ \sum_i x_i = 0.8$ and $\sum_i x_i^2 = (0.8)^2 $. It is then easy to
get both upper and lower limits to $V_8$ by considering the various
possible cases where $p$ values among the $x_i$'s are negative. [It
is easily seen that $p$ must be $ p \geq 1$.] Using then the
convexity inequality $ a_1^2 a_2^2 \cdots a_n^2 \leq [(\sum_i a_i^2)/n]^n$,
with $n = 8 -p$ and/or $n=p$, and minimizing over the average value
of the negative $x_i$'s, we get the double-sided inequality

\beq
 - 45 (0.8)^8/ 8^{4.5} \leq V_8/ (\SI_v)^8 \leq + 45 (0.8)^8/ 8^5
 \eeq
where we denote $\SI_v \equiv \sum_a v_a$. Numerically, this yields
for the quantity we are interested in:
\beq
 - 0.0391 \leq 60 V_8/ (\SI_v)^8 \leq + 0.0138
\eeq
These rather small allowed  values for $V_8$ are due to the fact that 
$V_8$ is a sum of products of 8 different $v$'s. By contrast, we expect 
the quantity $V_2^4/ (\SI_v)^8 $ in \Ref{j0vb} to typically  take 
significantly larger values because it contains products of only two 
or three different $v$'s. This suggestive (if not quite conclusive) 
argument makes it plausible that the difference $V_2^4 - 60 V_8$
will be always positive. Actually, the place where this difference
is most likely to become negative is in the vicinity of the
points where $(V_2)^4$ vanishes. This happens (normalizing again by  
$ \sum_a v_a = 1$) when one specific $v$ is very close to 1, and 
the other ones are near zero. [Note that the Kasner solution where 
$ (v_a) = (1,0,0,\cdots,0)$ is flat spacetime in disguise.] However, 
it is easy to see that, near such a point, $V_2^4 = \OC(v^4)$ 
while $V_8 = \OC(v^7)$. Therefore, near the points where $V_2^4$ is 
the smallest, the difference $V_2^4 -60 V_8$ remains positive. We 
performed some partial numerical investigations of the sign of the
difference $V_2^4 - 60 V_8$ and always found it to be positive.

We take all this as evidence that the total 8th order action \Ref{R4tot} 
is equivalent to a \textit{negative} contribution in the Hamiltonian, 
{\it i.e.} to an `inverted wall', or a `crevice'. As we said above, 
this is compatible with the $E_{10}$ expectation that there should 
be no bounce.

\section{Subleading terms: some examples}

We now wish to rephrase the results derived in section~4 in a Lie
algebraic setting, by giving more details concerning the subleading 
8th-order derivative corrections, and by illustrating their asymptotic 
behavior with some examples. More specifically, we will demonstrate 
explicitly how {\em all these terms can be associated either with (in 
general imaginary) roots of $E_{10}$, or, more generally, with elements 
of the $E_{10}$ root lattice.} Comparing the relevant highest roots with 
the corresponding highest weights in the tables of \cite{NiFi03} 
we can then read off to which $\gl$ multiplets they belong when 
they do correspond to actual roots.

In section~4, we estimated the various tensor structures in comparison
with the leading curvature terms $R \sim e^{2\SI}$. We have already
remarked that the subleading terms are smaller than the latter terms 
by exponential factors involving electric and symmetry wall forms,
which are linear combinations of the simple wall forms appearing 
in \Ref{simplewalls}. Above, we discussed a
coarse ordering of the effects of these wall forms by focussing
on the number of basic electric roots $\A_0$ entering them.
We shall now have a closer look at the various wall forms
associated with the different types of higher-order corrections,
and their relation to the root lattice of $E_{10}$.

From the results in Section 4 above, and from the basic fact that 
in the final expression all indices must be contracted in pairs, and 
thus appear {\em an even number of times}, we can see that
the subleading terms in the Hamiltonian constraint are all obtained 
by multiplying the leading Hamiltonian contribution $ e^{ 6\SI}$ 
(corresponding to the leading root $\A_{\rm leading} = - 3 \SI$) 
by an exponential factor of the form $\exp ({- 2 \sum n_i \A_i})$, 
with $n_i$ being some non-negative integers. Because of the presence 
of a factor two in the exponent (linked to the fact that all indices 
are finally contracted), we see that all the wall forms associated 
to the corrections can be written as $\exp ({- 2 w_{\rm subleading}})$, 
with subleading wall forms of the type \beq \label{cone}
w_{\rm subleading} = \A_{\rm leading} +\sum n_i \A_i
   \equiv - 3 \SI +\sum n_i \A_i
\eeq
We see from \Ref{cone} that the set of wall forms associated
with higher-order corrections is a subset of the root lattice of
$E_{10}$. To better understand this subset, it is convenient to
introduce at this point a geometric picture of this subset.
This is done in Figure 1.

\begin{figure}
\centering \includegraphics{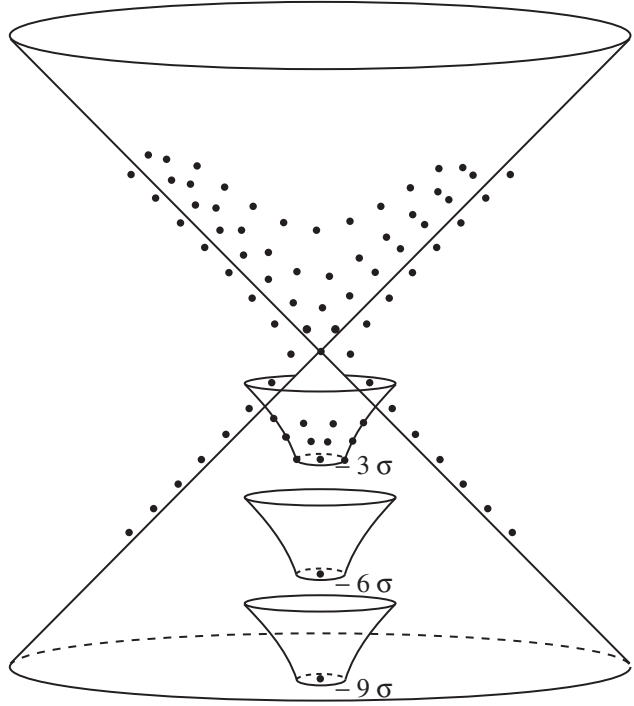}
\caption{\small Schematic representation of the roots of $E_{10}$
in the Lorentzian space of the Cartan degrees of freedom $\B$.
Positive roots are in the upper part of the figure, while
negative ones are in the lower part. The real roots are on the
outer (timelike) hyperboloid $\A^2 = +2$. The imaginary roots
are either null ($\A^2 = 0$) or timelike ($\A^2 = - 2 j$).
The basic root associated with the leading $R^4$ correction to the
supergravity action is the negative, timelike point $- 3 \SI$
indicated. From it stems the solid half-hyperboloid of roots associated 
with subleading contributions contained in $R^4$ as well as in the 
other 8th order derivative terms:
$R^2 (DF)^2$, $R (DF)^3$, $(DF)^4$, $\cdots, F^8$,
as well as the Chern-Simons contribution $A R^4$.
Similar half-hyperboloids stem from the roots $ - 3 k \SI$ predicted
to exist and to be associated to higher-order contributions
$ \sim R^7, R^{10}, \cdots$}
\label{Figure1}
\end{figure}

The reasoning above shows only that the wall forms are of the form 
\Ref{cone} for {\em some} non-negative integers $n_i$. 
The tools we used above are too coarse to determine whether 
{\em all} values of the integers $n_i$ are realized, or whether 
there are restrictions of any kind on the values of the $n_i$'s. 
In view of the complicated structure of the higher-order 
corrections, which as we pointed out, is not known in an 
algebraically completely reduced form, especially for the 
crucial next-to-leading terms $R^2 (DF)^2$ and $R^3 F^2$),
we have made no attempt at an exhaustive analysis as to whether
the integers $n_i$ are indeed subject to restrictions. 
Instead, we have proceeded `experimentally' by studying in detail 
some specific, but typical, terms that we could check to be indeed 
present among the available expressions for various higher-order 
corrections. In all the cases we have checked, we did find a 
rather remarkable pattern: the vector in root space joining 
the `leading ' $R^4$ root $\A_{\rm leading} = - 3 \SI$ to any subleading 
higher-order wall form $w_{\rm subleading}$, {\em i.e.} the quantity
\beq \label{vector}
\zeta := w_{\rm subleading} - \A_{\rm leading} = \sum n_i \A_i
\eeq
was `experimentally' found to satisfy the restriction
\beq \label{hyper}
\zeta^2 \equiv \left(\sum n_i \A_i\right)^2 \leq 2
\eeq
in all cases. Below, we will refer to $\zeta$ as the `relative vector'.
This leads us to formulate the following

\vspace*{0.2cm}\noindent
{\bf{Conjecture:}} {\it The relation \Ref{hyper} holds for all 
kinematically allowed combinations, and thereby encodes information 
about the algebraic structure of the 8th order corrections.}

\vspace*{0.2cm}

We shall illustrate below with some examples the interrelation between 
this conjecture and the properties of the $E_{10}$ root lattice on
the one hand, and certain `kinematical cancellations', as they follow 
from the kinematical structure of the known 8th order correction terms, 
on the other hand. However, we will leave a systematic investigation of 
this aspect to future work. 
 
Geometrically, the restriction \Ref{hyper} (together with $n_i \geq 0$)
means that the relative position vector $\zeta$, Eq. \Ref{vector},
belongs to the set of positive roots, $\Delta_{+}$.
In other words, if we assume that this restriction is indeed true
for all terms, we can geometrically describe the set of wall forms 
generated by higher-order corrections as a solid half-hyperboloid, 
which is congruent to $\Delta_{+}$, and with the leading $R^4$
root $\A_{\rm leading} = - 3 \SI$ as its basis, see Figure 1.
From this picture there seems to be no evident upper limit on the 
height $\sum n_i$ of the relative vector \Ref{vector}. We initially
thought that the wall forms $w_{\rm subleading}$ might be
constrained to lie inside the root diagram {\it stricto sensu}
(instead of lying anywhere in the root lattice),
which would have implied the further restriction
$ (w_{\rm subleading})^2 \leq 2$. However, as we shall see
below, we found wall forms lying high enough in the solid
half-hyperboloid \Ref{hyper} to extend beyond the outer
hyperboloid $w^2 = 2$ corresponding to real roots.
More precisely, by adding simple roots of $A_9$ one generates
$\gl$ weight diagrams starting at a certain highest weight inside
this solid half-hyperboloid. Many of the $\gl$ weight diagrams are
such that only a subset of the weights are actually roots of $E_{10}$,
while the weight diagram itself extends beyond the hyperboloid $w^2 =2$.
We emphasize that there is no a priori inconsistency or incompatibility 
in this feature: indeed, starting from a Hamiltonian containing only 
roots as wall forms, any change of parametrization can generate arbitrary 
combinations of the roots, and therefore arbitrary elements of the root 
lattice. 

Let us now substantiate our conjecture \Ref{hyper} by giving examples 
of wall forms associated by various higher-order contributions.
We first consider the Chern Simons term in \Ref{i2}, which is
proportional to
\bea
&& \E^{ABCD_1 \dots D_8} A_{ABC}
{R_{D_1D_2}}^{E_1E_2} \cdots {R_{D_7 D_8}}^{E_7 E_8}
t^{(8)}_{E_1 \cdots E_8} \nn
&& \qquad\qquad\equiv 
 \E_{11} A_3 \left[ 24 \, \tr R^4 - 6\, (\tr R^2)^2\right] 
\eea
Here the tensor $t_8 $ is, by virtue of \Ref{t8}, equivalent to the 
combination of traces appearing in \Ref{i2}. Though the above 
contribution is also quartic in the curvature, and contains the 
spatial components $A_{abc}$ of the three-form which simply freeze near
the singularity, it is actually subleading w.r.t. the leading curvature 
contribution $ \sim E R^4 \sim e^{6 \SI}$. The reason for this is the 
interplay of the  $\E$ tensor, which constrains the first pairs of 
indices $[D_1D_2] \cdots [D_7 D_8]$ on the curvature tensors to be all 
different, with the effect of the $t_8$ tensor, which contains Kronecker 
deltas and thereby obliges the remaining indices on the curvature tensors 
to coincide pairwise. As a consequence, it is not possible to have 
only leading curvature components of the type $R_{0a0a}$ or $R_{abab}$. 
A typical contribution is \footnote{In the remainder, we will no longer
distinguish between the Riemann and the Weyl tensor.}
\beq
E A_{89\, 10} R_{0101} R_{0123} R_{0145} R_{0167} \sim
e^{3\SI + 3 \B^1 + \B^2 - \B^3 + \B^4 - \B^5 + \B^6 - \B^7 - \B^8 -
    \B^9 - \B^{10}}
\eeq
where we used \Ref{R0abc}. After reshuffling some indices, we obtain 
the wall form
\beq
w = - (3,2,2,2,1,1,1,1,1,1) \quad\Longleftrightarrow\quad
w = - [ 5\, ; \, 2, 5, 8, 6, 5, 4, 3, 2, 1 ]
\eeq
This is a root because $w^2 =2$, with $\ell = -5$. The corresponding 
relative vector
\beq
\zeta = w - \A_{\rm leading}
\equiv w + 3 \SI = (0,1,1,1,2,2,2,2,2,2)
\eeq
satisfies $ (w + 3 \SI)^2 = + 2$, in agreement with \Ref{hyper}.
We thus see that this Chern Simons wall form is so much higher than 
the basic $R^4$ root $ - 3 \SI$ that the associated wall form lies 
even beyond the interior of the past CSA lightcone and belongs to 
the timelike hyperboloid $\A^2 = +2$.

Next we consider the subleading curvature terms. Let us first recall
that the kinematic structure of the $C^4$ terms in \Ref{C4} eliminates 
certain combinations; for instance, from \Ref{j0vb} it is obvious that
there are no contributions of the type $C_{abab}^4$ (no summation on
indices $a,b$).
As is evident from \Ref{Rabcd2}, the curvature terms coming from 
$R^{(2)}_{abcd}$ and the above estimates are strongly suppressed.
Furthermore, not all of them produce actual roots, as we shall see
presently.

Obviously, from \Ref{Rabcd2}, there are many such terms, which we 
can group into certain `permutation multiplets'. A first set of terms 
corresponds to terms of type
\bea
E R^{(2)}_{abcd} R^{(2)}_{abef} R^{(2)}_{cegh} R^{(2)}_{dfgh} &\sim&
    e^{-2\SI} \cdot e^{\B^a + \B^b + \B^c - \B^d} \cdot 
    e^{\B^a + \B^b + \B^e + \B^f - 2\B^i} \nn
&& \cdot 
e^{\B^c +\B^e +\B^g +\B^h -2\B^j} \cdot e^{\B^d +\B^f +\B^g +\B^h -2\B^k}
\eea
We can write this result schematically as
\beq
E \left[ R^{(2)} \right]^4 \sim
e^{- 2\SI}\cdot e^{2(\B^a + \B^b + \B^c + \B^e + \B^f + \B^g 
     + \B^h - \B^i -\B^j - \B^k)}
\eeq 
Taking the indices to be all different, and choosing $(i,j,k)= (8,9,10)$, 
this gives
\beq\label{R24}
w = -(0,0,0,0,0,0,0,2,2,2)  \quad\Longleftrightarrow\quad
w= - [2 \, ; \, 2 , 4 , 6 , 6 , 6 , 6 , 6 , 4 , 2 ]
\eeq
which is clearly {\em not} a root as $w^2= + 8$. Nevertheless, the 
relative vector
\beq
\zeta = w + 3\SI = (3,3,3,3,3,3,3,1,1,1)
\eeq
obeys $\zeta^2 = 2$ and therefore lies on the boundary of the half-hyperboloid
in agreement with our conjecture. A second set of terms derives from
\beq
E \left[ R^{(2)} \right]^4 \sim
e^{- 2\SI}\cdot e^{+ 2(\B^a + \B^b + \B^c + \B^d + \B^e + \B^f -\B^g  -\B^h)}
\eeq 
and differs from the one above by one symmetry wall. Again taking 
all indices different and choosing them appropriately, we get
\beq
w = -(0,0,0,0,0,0,1,1,2,2)  \quad\Longleftrightarrow\quad
w= - [2 \, ; \, 2 , 4 , 5 , 6 , 6 , 6 , 6 , 4 , 2 ]
\eeq
which is again not a root as $w^2= + 6$. The relative vector
\beq
\zeta = w + 3\SI = (3,3,3,3,3,3,2,2,1,1)
\eeq
is lightlike, $\zeta^2 = 0$, and lies within the half-hyperboloid.
A third set of terms comes from
\beq
E \left[ R^{(2)} \right]^4 \sim
e^{- 2\SI}\cdot e^{+ 2(\B^a + \B^b + \B^c + \B^e + \B^g - \B^h)}
\eeq 
and gives rise to, choosing values for the indices conveniently,
\beq
w = -(0,0,0,0,0,1,1,1,1,2)  \quad\Longleftrightarrow\quad
w= - [2 \, ; \, 2 , 4 , 6 , 6 , 6 , 5 , 4 , 3 , 2 ]
\eeq
with $w^2 = +4$, and 
\beq
\zeta = (3,3,3,3,3,2,2,2,2,1)
\eeq
with $ \zeta^2 = - 2$. Finally, there is a fourth set with
\beq
w= -(0,0,0,0,1,1,1,1,1,1) \quad\Longleftrightarrow\quad
w= - [2 \, ; \, 2 , 4 , 6 , 6 , 5 , 4 , 3 , 2 , 1 ]
\eeq
which now {\em is} a root because $w^2 = 2$, and
\beq
\zeta = (3,3,3,3,2,2,2,2,2,2) \quad \Rightarrow \quad \zeta^2 = -4
\eeq
again confirming the conjecture. The maximal height for any of these 
wall forms $w$ is $- 44$ for \Ref{R24}, well above the height $- 115$ 
computed above for the leading term, see Figure 1. We observe that all 
of the terms displayed above differ from one another either by simple 
permutations of the wall form components, or by the addition of 
symmetry roots.

Similar comments apply to the curvature terms with one or more temporal
indices. From \Ref{C4}, we find, for instance, that terms 
$\propto (C_{0abc})^4$ or $(C_{0abc})^2 (C_{0ade})^2$, with the 
indices $a,b,c,d,e$ all different (and no summation on repeated
indices!), cancel between the two contributions in \Ref{C4}.
Remarkably, these terms are precisely of a form which is disallowed by 
our conjecture above: for instance, if the term
\beq
E (R_{01ab})^2 (R_{01cd})^2 \sim e^{-2\SI} \cdot 
\left(e^{\SI + \B^1 + \B^a - \B^b}\right)^2 
\left(e^{\SI + \B^1 + \B^c - \B^d}\right)^2 
\eeq
with $a,b,c,d$ all different, had contributed instead of cancelling, it
would have yielded (choosing indices conveniently)
\beq 
w = - (3,2,2,1,1,1,1,1,0,0) \quad\Rightarrow\quad
\zeta = (0,1,1,2,2,2,2,2,3,3)
\eeq
whence $\zeta^2= +4$, in violation of \Ref{hyper}.

A set of terms which {\em does} contribute, by inspection of \Ref{C4}, is
\beq
E (R_{01 ab})^2 (R_{0101})^2 \sim
e^{4\SI - 2\B^1 + 2\B^a + 2\B^{b}}
\eeq
Choosing $(a,b)= (9,10)$, we get the maximal height contribution
\beq
w = - (1,2,2,2,2,2,2,2,3,3) \quad\Longleftrightarrow\quad
w = - [7\, ; \, 6, 11, 16 , 14 , 12 , 10 , 8 , 6, 3]
\eeq
at level $\ell = -7$, of height $- 93 > - 115$, and obeying $w^2 = -2$. 
The corresponding relative vector is
\beq
\zeta = (2,1,1,1,1,1,1,1,0,0) \quad\Rightarrow\quad \zeta^2 = 2
\eeq
and therefore on the boundary of the half-hyperboloid. From the tables
of \cite{NiFi03} we deduce that the set of roots corresponding to 
$w$ belongs to the $\gl$ representation $(010000001)$ with outer 
multiplicity $\mu = 1$.

There are numerous mixed terms involving the curvature and the 4-form 
field strength, which we again illustrate with some examples (we have 
checked from \cite{DS} that the structures displayed below actually 
do appear). For instance, the term $E(D_0 F_{0123})^2 (D_0 F_{0456})^2$ 
leads to the $\ell = 8$ root
\beq
w = -(3,3,3,3,2,2,2,2,2,2) \quad\Longleftrightarrow\quad
w = - [8\,; \, 5, 10, 15 , 12 , 10 , 8 , 6 , 4, 2]
\eeq
with $w^2 = -4$, and relative vector
\beq
\zeta = (0,0,0,0,1,1,1,1,1,1) \quad\Rightarrow\quad \zeta^2 = 2
\eeq
At level $\ell = -8$, it corresponds to the $\gl$ representation
$(0001000000)$ with $\mu =2$. The highest root in the multiplet is
\beq
w = -(2,2,2,2,2,2,3,3,3,3) \quad\Longleftrightarrow\quad
w = - [8\, ; \, 6, 12 , 18 , 16 , 14 , 12 , 9, 6 , 3]
\eeq
of height $-106 > - 115$, well above the height of the leading 
$R^4$ singlet.

From the $C(DF)^3$ terms given in \cite{DS} we read off the leading
term which is $E D_0 F_{3456} D_0 F_{789\, 10} D_a F_{a012} C_{0b0b}$.
The associated root is
\beq
 w = -(2,2,2,2,2,2,2,2,1,1) = - [6\, ; \, 4, 8, 12 , 10 , 8 , 6 , 4, 2 , 1]
\eeq
with $w^2 = -2$, and
\beq
\zeta = (1,1,1,1,1,1,1,1,2,2) \quad\Rightarrow\quad \zeta^2 = 0
\eeq
The relevant $\gl$ representation at level $\ell =6$ is
$(000000010)$ with $\mu =1$. The maximal height in the representation
is again above $- 115$.

Our final example is the combination $E(D_0 F_{0123})^2 (C_{0a0a})^2$. 
It is associated to the $\ell =9$ root
\beq
w = -(3,3,3,3,3,3,3,2,2,2) = - [9\, ; \, 6, 12, 18 , 15 , 12 , 9 , 6 , 4, 2]
\eeq
with relative vector
\beq
\zeta = (0,0,0,0,0,0,0,1,1,1) \quad\Rightarrow\quad \zeta^2 =2
\eeq
Indeed, in agreement with the result \Ref{DF} above, we see that  
$\zeta$ is just an electric root. On the other hand, $w^2 = -6$, and 
$w$ corresponds to the $\sl$ representation $(000000100)$ with $\mu =4$. 
The highest root in the multiplet is 
\beq
w = -(2,2,2,3,3,3,3,3,3,3) \quad\Longleftrightarrow\quad
\A = - [9\, ;\, 7, 14, 21 , 18 , 15 , 12 , 9, 6 , 3]
\eeq
of height $- 114 > -115$, barely above the height of the leading 
singlet $C^4$ contribution.

\section{Discussion}

Our analysis of higher order corrections to M Theory provides
further evidence for the validity of the conjecture \cite{DaHeNi02}
that the classical (bosonic) dynamics of M theory is `dual' to a 
one-dimensional $\SI$-model on the infinite dimensional coset
space $E_{10}/K(E_{10})$. In particular, we find it remarkable
that the leading wall form associated with the known $R^4$ corrections,
namely the {\em permutation singlet} $w_4 = - 3 \SI$ does match
with a root of $E_{10}$ whose corresponding generator is a
{\em $\gl$ singlet}. This compatibility between a gravity
structure and a Kac-Moody algebra one was not a priori guaranteed,
and can be viewed as a deep confirmation of the hidden role of
$E_{10}$ in M theory. Indeed, as a foil, let us consider the simple
case of pure gravity in any spacetime dimension $D \equiv d+1$.
The study of the corresponding cosmological billiard has found
that one should associate to pure gravity the Kac-Moody algebra 
$AE_d$ \cite{DaHeJuNi01}.
Now, for pure (bosonic) gravity, one generally expects that the 
first higher-order corrections will be $\sim (R_{\mu\nu\rho\sigma})^2$. 
However, by using Eq. \Ref{wRN} such terms quadratic in curvature
correspond to the wall form 
$w_2 = - \, \sigma$ whose squared length is $(w_2)^2 = - \, d/(d-1)$. 
As the latter squared length is never an integer \footnote{For
$d \geq 3$. Actually, we should restrict ourselves to $d \geq 4$ 
as all $R^2$ terms are known to be on-shell trivial in 
$d=3$ because of the topological nature of the $R^2$ Euler-Gauss-Bonnet 
density.}), we conclude that $w_2$ {\em never corresponds to a 
root of $AE_d$}. Let us then consider the problem of determining 
which values, if any, of the non-linearity order $N$, for corrections 
$\sim R^N$, might be compatible with the algebraic structure of $AE_d$. 
For instance, let us consider the case $d=3$ corresponding to the 
usual $(3+1)$-dimensional Einstein gravity. By using again Eq. \Ref{wRN},
 one gets the wall form $w_N = - \, (N-1) \, \sigma$ whose squared 
length in $AE_d$ is $- \, \frac{d}{d-1} \, (N-1)^2$; i.e. 
$- \, \frac{3}{2} \, (N-1)^2$ when $d=3$. One then concludes that one 
needs $N = 2k+1 = 3,5,7, \ldots$ The lowest candidate for compatibility 
with $AE_3$ is then $\sim R^3$. The corresponding wall form is 
$w_3 = - \, 2 \, \sigma = - \, 2 \, (\beta^1 + \beta^2 + \beta^3)$. 
Its squared length is $(w_3)^2 = - \, 6$, and $w_3$ is easily seen 
to correspond to a root $\alpha$ of $AE_3$ of level $\ell = 3$ w.r.t. 
the $SL(3)$ subalgebra \cite{DaHeNi03}. However, as in the case discussed 
above for $E_{10}$, the wall form 
$w_3 = - \, 2 \, (\beta^1 + \beta^2 + \beta^3)$ is invariant under 
permutations of the spatial indices. Therefore, for the conjectured 
correspondence between $(3+1)$-dimensional Einstein gravity and 
$AE_3$ to hold, the wall form $w_3$ should correspond to a root of 
$AE_3$ which parametrizes a {\em singlet} of $SL(3)$. However, by using 
the results given in Eq. (8.30) of \cite{DaHeNi03} one sees that 
{\em there is no singlet representation} of $AE_3$ at level $\ell = 3$. 
This negative result exemplifies that the compatibility found above 
between $\sim R^4$ corrections and the presence of `singlet roots' 
of $E_{10}$ is rather non trivial, and could well have failed to hold.

Our work has also provided evidence for the `no bounce' 
behavior of big crunches in M theory, as naively expected from the 
dual dynamics, {\it i.e.} the global structure of null geodesics 
on $E_{10}/K(E_{10})$. If we admit the validity of this
conjecture, what conclusions can we draw for big crunches in M theory?
The `dual' description is an infinite affine length null geodesic going 
towards $\B\rightarrow\infty$. This suggests that the quantization of
the  $E_{10}/K(E_{10})$ model will exhibit no information loss at the
big crunch. On the other hand, the dictionary of \cite{DaHeNi02,DaNi04}
between the supergravity description and the coset description is 
defined only in the quasi-classical regime $T\gg T_P$. BY contrast, the 
infinite future of the null geodesic motion corresponds to the regime
$T\ll T_P$. The fact that the dictionary between the two descriptions
becomes ill-defined in this limit (and that we find no evidence for a bounce)
suggests that the infinite `affine life' near the singularity can only
be described in the coset variables. This situation is somewhat
reminiscent  of the recent results of \cite{HH} based on an AdS/CFT 
analysis of certain cosmological singularities in anti-deSitter solutions
of supergravity.\footnote{The fact that the dual $E_{10}$ coset
picture emerges in full only in the `strongly coupled limit' $T\ll T_P$,
while the gravity picture corresponds to the `weakly coupled limit'
$T\gg T_P$ is also reminiscent of what happens in the AdS/CFT duality.}
 [For a contrasting suggestion based, similarly to
the $E_{10}$ coset model, on geodesic motion in auxiliary Lorentz
spaces, see refs. \cite{TW,RT05}].

However, we feel that at this stage of development of the  $E_{10}/K(E_{10})$
conjecture, speculating on the ultimate quantum fate of big crunches is
premature. Many technically challenging tasks  remain before one can
seriously consider the eventual physical consequences of the dual
$E_{10}$ picture. First, one should extend the dictionary to prove the
heretofore unseen roots between height 29 and height 115 do match in
the two descriptions.
Second, it would be interesting to explore in more detail the
validity of the conjecture \Ref{hyper}, which was `experimentally'
observed to hold in quite a few specific cases. If our conjecture could
be fully verified for the {\em known} 8-th order corrections, this would
open the tantalizing possibility that it still holds for higher
corrections $\sim (R + DF + F^2)^7, (R + DF + F^2)^{10}, \dots$.
It would then provide a strong constraint on the algebraic structure
of these corrections. As these corrections are very difficult to
obtain by conventional methods, the hidden $E_{10}$ structure might
be of great help in pinning down their structure.

Finally, among other pressing issues let us also mention: the role 
of fermions, the consequences of compactifying eleven-dimensional
spacetime (which is expected to reduce the continuous symmetry
$E_{10}(\Rn)$ to the discrete symmetry $E_{10}(\Zn)$), and the effect 
of quantizing the coset model (see, in this respect, also the remarks 
in the Appendix below). It would furthermore be interesting to explore 
the link, if any, between the non-compact T-duality symmetries of string 
theories in presence of Killing vectors (which have been shown to survive 
the addition of higher-order terms \cite{Meissner:1996sa}) and the 
conjectured $E_{10}$ symmetry.


\vglue 8mm
\noindent{\bf Acknowledgements}

\noindent We are grateful to Nathalie Deruelle, Thomas Fischbacher,
Gary Horowitz, Victor
Kac, Axel Kleinschmidt, Kasper Peeters, Jan Plefka and Arkady Tseytlin 
for useful exchanges of information. Special thanks go to Ofer Gabber 
for suggesting a nice way to bracket the values of the term  $V_8$. 
We wish also to thank Stanley Deser for many enlightening exchanges 
over the years, and the organizers of the Deserfest where this work 
was initiated. We are grateful to Marie-Claude Vergne for
preparing the figure.
H.N. thanks the Institut des Hautes Etudes Scientifiques
for hospitality during the maturation of this work, while T.D. thanks 
the Albert Einstein Institut for its hospitality during its completion.

\newpage

\section*{Appendix}
\appendix
\subsection*{Geodesic deviation and sectional 
curvature on $\mbox{$\boldmath{E_{10} / K(E_{10})}$}$}

As shown in \cite{DaHeNi02,DaHeNi03}, geodesic motion on the coset 
space $E_{10} / K(E_{10})$ is formally integrable in the sense that 
one can exhibit as many constants of motion as degrees of freedom 
(i.e. a doubly infinite tower ${\mathcal J}$) and, moreover, that one 
can formally write a generic solution of the geodesic equation in terms 
of the constants of motion ${\mathcal J}$. We wish, however, to emphasize here 
that, even if formally integrable, geodesic motion on $E_{10} / K(E_{10})$ 
is at the same time strongly chaotic in the sense that it is extremely 
sensitive to small deviations in the initial conditions. Let us, indeed, 
consider the `geodesic deviation equation' which governs the deviation 
$\mbox{\boldmath$\xi$}$ between two neighbouring geodesics. For simplicity, 
we focus on the neighbourhood of a base geodesic lying entirely in the CSA 
(corresponding to a simple Kasner solution), i.e. where only $\beta^a$ 
and $\pi_a$ are excited. This solution is simply described by 
$\beta^a = v^a \, t + \beta_0^a$ where $t$ is an affine parameter, 
$\beta_0^a$ some constants and where the `velocity' $v^a$ is null: 
$G_{ab} \, v^a \, v^b = 0$. The other (`off diagonal')
variables, $ \nu, p$, are all zero. The geodesic deviation equation
governing the affine-parameter evolution of the deviation 
vector $\mbox{\boldmath$\xi$}$ connecting this geodesic 
to a neighbouring one can be obtained in two different ways: 
either by `moving' the base geodesic $\beta^a = v^a \, t + \beta_0^a$ 
by a generic element in the Lie algebra of $E_{10}$, or by computing the 
appropriate components of the sectional curvature of the symmetric 
space $E_{10} / K(E_{10})$ and by writing the general geodesic-deviation 
(Jacobi) equation. We have checked that both methods give the same result. 
If we decompose the deviation vector $\mbox{\boldmath$\xi$}$ in its various 
components : 
$$
\mbox{\boldmath$\xi$} = \xi^a \, h_a + {\sum}_{\alpha , s} \ 
\xi_{(s)}^{\alpha} (E_{\alpha}^{(s)} + F_{\alpha}^{(s)}) / \sqrt 2, 
$$
where $h_a$, $a = 1 , \ldots , 10$, is a basis of the CSA, and where 
$\xi_{(s)}^{\alpha}$ denote the components along the non-CSA coset directions.
The Jacobi equation yields $\ddot{\xi}^a = 0$, corresponding to a vanishing 
sectional curvature in the $vv'$ two-planes (where $v = v^a \, h_a$ is 
the vector tangent to the base geodesic and $v' = v'^a \, h_a$ a generic 
direction in the CSA), and 
$\ddot{\xi}_{(s)}^{\alpha} = + \, (\alpha (v))^2 \, \xi_{(s)}^{\alpha}$, 
corresponding to a sectional curvature 
$$
\R(v , E_{\alpha}^+ , v , E_{\alpha}^+) = - \, (\alpha (v))^2
$$ 
where $E_{\alpha}^+ \equiv (E_{\alpha} + F_{\alpha}) / \sqrt 2$ is a 
unit-norm vector in the `transverse' coset direction associated to the 
root $\alpha$. [For simplicity, we suppressed the degeneracy 
index $s$ on $E_{\alpha}^{(s)}$ and $F_{\alpha}^{(s)}$.] Here 
$\alpha (v) \equiv \alpha_a \, v^a$ denotes the value of the linear form 
$\alpha$ acting on the vector $v = v^a \, h_a$ tangent to the geodesic. 
The point we wish to emphasize here is two-fold: $(i)$ the sectional curvature
$\R (v , E_{\alpha}^+ , v , E_{\alpha}^+)$ ~\footnote{As $E_{10} / K(E_{10})$ 
is a homogeneous space the sectional curvature components computed here 
should be the same in the neighbourhood of an arbitrary geodesic.} is 
{\em negative} (as for geodesics on hyperbolic space), and $(ii)$ this 
sectional curvature {\em decreases without limit towards $ - \infty$} as 
the height of the root increases to infinity. This result shows that 
small deviations from the geodesic exponentially increase with affine 
length in a manner which becomes faster and faster as we consider 
components of the deviation in transverse directions corresponding 
to higher and higher roots. This indicates that geodesics on 
$E_{10} / K(E_{10})$ are more and more exponentially sensitive to 
initial conditions as one examines them in directions of increasing 
heights. This property might play an important role for the quantization 
of the geodesic motion on $E_{10} / K(E_{10})$. Indeed, we recall that 
quantum motion for a scalar particle on curved space involve an ambiguous 
coupling to the scalar curvature, which appears to be given by a 
formally infinite series for $E_{10} / K(E_{10})$. This suggests that 
it is important to include fermions in the $E_{10}$ coset model before 
drawing conclusions about the strong sensitivity of classical geodesic 
motion on initial conditions.

\newpage


%
\end{document}